\documentclass[useAMS,usenatbib,fleqn]{mnras}
\usepackage{epsfig}
\usepackage{deluxetable} 
\usepackage{url}
\usepackage{rotating}
\usepackage{amsmath}
\usepackage{graphicx}
\usepackage{setspace}
\usepackage{psfig}
\usepackage{natbib}
\usepackage{float}
\usepackage{placeins}
\usepackage{booktabs}

\title[NICER, NuSTAR and Swift observations of BL Lac]{NICER, NuSTAR, and Swift follow-up observations of the $\gamma$-ray flaring blazar BL Lacertae in 2020 August--October}
\author[D'Ammando F.]{F. D'Ammando$^{1}$\thanks{E-mail: dammando@ira.inaf.it} \\
$^{1}$INAF - Istituto di Radioastronomia, Via Gobetti 101, I-40129 Bologna, Italy\\
}

\begin{document}

\date{Accepted. Received; in original form}

\maketitle

\label{firstpage}

\begin{abstract}

During a period of strong $\gamma$-ray flaring activity from BL Lacertae, we organized {\em Swift}, {\em Neutron star Interior Composition Explorer (NICER)}, and {\em Nuclear Spectroscopic Telescope Array (NuSTAR)} follow-up observations. The source has been monitored by {\em Swift}-XRT (X-ray Telescope)  between 2020 August 11 and October 16, showing a variability amplitude of 65, with a flux varying between 1.0$\times$10$^{-11}$ and 65.3$\times$10$^{-11}$ erg\,cm$^{-2}$ s$^{-1}$. On 2020 October 6, {\em Swift}-XRT has observed the source during its historical maximum X-ray flux. A softer-when-brighter behaviour has been observed by XRT, suggesting an increasing importance of the synchrotron emission in the X-ray part of the spectrum covered by XRT during this bright state. Rapid variability in soft X-rays has been observed with both the {\em Swift}-XRT and {\em NICER} observations with a minimum variability time-scale of 60 and 240 s, and a doubling time-scale of 274 and 1008 s, respectively, suggesting very compact emitting regions (1.1$\times$10$^{14}$ and 4.0$\times$10$^{14}$ cm). At hard X-rays, a minimum variability time-scale of $\sim$5.5 ks has been observed by {\em NuSTAR}. We report the first simultaneous {\em NICER} and {\em NuSTAR} observations of BL Lacertae during 2020 October 11--12. The joint {\em NICER} and {\em NuSTAR} spectra are well fitted by a broken power law with a significant difference of the photon index below (2.10) and above (1.60) an energy break at $\sim$2.7 keV, indicating the presence of two different emission components (i.e. synchrotron and inverse Compton) in the broad band X-ray spectrum. Leaving the total hydrogen column density towards BL Lacertae free to vary, a value of N$_{\rm\,H,tot}$ = (2.58 $\pm$ 0.09) $\times$10$^{21}$ cm$^{-2}$ has been estimated.

\end{abstract}

\begin{keywords}
radiation mechanisms: non-thermal; galaxies: active; galaxies: jets; X-rays: galaxies 
\end{keywords}

\section{Introduction}

The extragalactic $\gamma$-ray sky is dominated by blazars, a class of radio-loud active galactic nuclei (AGN) in which one of the two relativistic jets points in the direction of the Earth. The  spectral energy distribution (SED) of blazars shows a non-thermal continuum from radio to $\gamma$-rays, characterized by two distinct components: one peaking in infrared-to-X-rays and associated with synchrotron emission by e$^{\pm}$, and one peaking in the $\gamma$-ray energy range and associated, in leptonic models, with inverse Compton (IC) scattering between the e$^{\pm}$ and a soft photon field (which can be their own synchrotron radiation, or an external photon field such as the emission from the broad-line region, the torus, or the disc). A common feature of all blazars is to show strong flux and spectral variability at all wavelengths and on a variety of time-scales, from minutes to years \citep[e.g.,][]{wagner95, ulrich97}. The variability time-scale of the emission can give constraints on the size of the emitting region \citep[e.g.,][]{tavecchio98, tavecchio10}. 

Blazars are classified into BL Lac objects and flat-spectrum radio quasars (FSRQ) according to the presence or absence of broad emission lines \citep[equivalent width EW $>$ 5 \AA; e.g.,][]{stickel91} in their optical spectrum, respectively. A further classification is based on the synchrotron peak frequency: While FSRQ show generally a low peak frequency (in infrared), in BL Lac objects the peak frequency ranges from radio to X-rays and are further classified as low/intermediate/high-frequency-peaked BL Lacs \citep[LBL, IBL, and HBL; with peak frequency lower than 10$^{14}$ Hz, between 10$^{14}$ and 10$^{15}$ Hz, and above 10$^{15}$ Hz, respectively; see e.g.][]{padovani95}\footnote{A similar classification (low/intermediate/high-synchrotron-peaked blazar) has been introduced in \citet{abdo10} not distinguishing BL Lac and FSRQ and making an explicit reference to the synchrotron origin of the ﬁrst bump of the SED.}. Therefore, observations in the X-ray energy range can cover different parts of the SED depending on the type of objects: the peak of the synchrotron component in HBL, the valley between synchrotron and IC components in IBL, the rising part of the IC component in LBL and FSRQ. 

BL Lacertae, a blazar at redshift $z$ = 0.069 \citep{miller97}, has been historically defined as the prototype of the BL Lac objects. According to its synchrotron peak frequency, it has been classified as an LBL \citep{nilsson18} or IBL \citep{ajello20}. Optical spectra of BL Lacertae have shown broad H$\alpha$ and H$\beta$ lines \citep{vermeulen95} that vary in flux \citep{capetti10}. The discovery of broad emission lines, although weak, in BL Lac seems to indicate that BL Lac itself can be more similar to FSRQ than to less luminous BL Lac objects. BL Lacertae has been studied intensively since its discovery in $\gamma$-rays by the Energetic Gamma Ray Experiment Telescope (EGRET) onboard the Compton Gamma Ray Observatory satellite \citep{catanese97} and has been the target of several observational campaigns in the last two decades from radio to very high energies \citep[e.g.,][and the references therein]{villata02, villata09, bach06, raiteri10, raiteri13, wierzcholska15, wehrle16, abeysekara18, acciari19}.

After a strong $\gamma$-ray flaring activity detected by the Large Area Telescope (LAT) onboard the {\em Fermi Gamma-Ray Space telescope} \citep{ojha20} and the MAGIC telescopes \citep{blanch20a} on 2020 August 19, BL Lacertae remains in a high activity in the following weeks and {\em Swift} observations performed on 2020 October 5 and 6 found the source at the historical maximum level in X-rays and in one of the brightest states observed in optical and ultraviolet \citep[UV][]{dammando20a,dammando20b}. On 2020 October 6, the source reached the second highest daily averaged $\gamma$-ray flux observed from {\em Fermi}-LAT so far \citep{mereu20}. Following the historical maximum flux observed from optical to $\gamma$-rays, simultaneous {\em NICER} and {\em NuSTAR} follow-up observations have been requested. Preliminary results are presented in \citet{dammando20c}. Results about the multi-frequency campaign over the entire electromagnetic spectrum will be presented in a separate publication. In this paper, we focus on the {\em Swift}, {\em Neutron star Interior Composition Explorer (NICER)} and {\em Nuclear Spectroscopic Telescope Array (NuSTAR)} data of BL Lacertae collected during 2020 August--October. We present the observations and data reduction in Section~\ref{reduction}. The results of analysis of the data collected by the three satellites separately and the joint fit of the {\em NICER} and {\em NuSTAR} X-ray data are shown in Sections~\ref{results} and~\ref{joint}, respectively. In Section~\ref{summary}, we discuss and summarize our results.

Unless stated otherwise, uncertainties correspond to 90 per cent confidence limits on one parameter of interest ($\Delta\chi^{2}$ = 2.7). The photon indices are parameterized as $N(E) \propto E^{-\Gamma}$ with $\Gamma = \alpha +1$ ($\alpha$ is the spectral index). Throughout this paper, we assume the following cosmology: $H_{0} = 71\; {\rm km \; s^{-1} \; Mpc^{-1}}$, $\Omega_{\rm M} = 0.27$, and $\Omega_{\rm \Lambda} = 0.73$ in a flat Universe \citep{planck16}. At the redshift of the source the luminosity distance D$_{L}$ is 307 Mpc.

\begin{figure*}
\begin{center}
\includegraphics[width=0.75\textwidth]{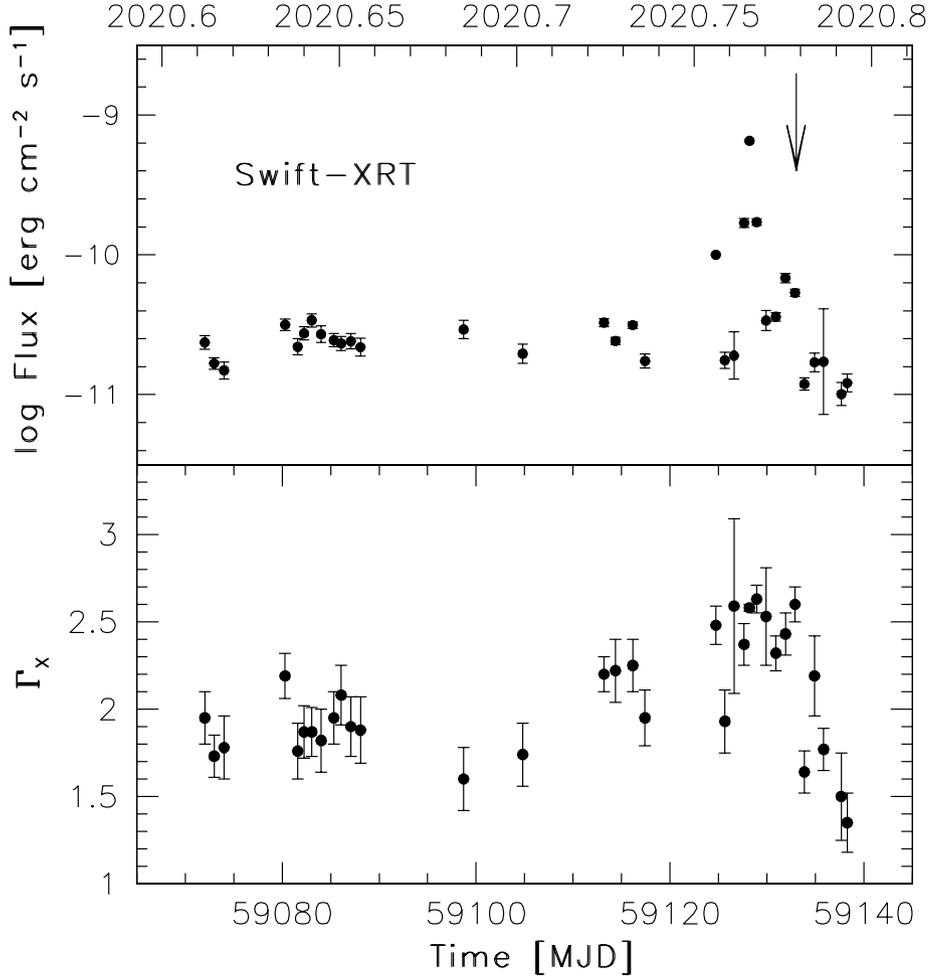}
\caption{{\em Swift}-XRT light curve of BL Lacertae in terms of flux (top panel) and photon index (bottom panel). The arrow indicates the time of the {\em NICER} and {\em NuSTAR} observations.}
\label{XRT_lc}
\end{center}
\end{figure*}

\section{Observations and data reduction}\label{reduction}

\subsection{{\em Swift} observations}\label{Swift_obs}

The {\em Neil Gehrels Swift observatory} satellite \citep{gehrels04} carried out 33 observations of BL Lacertae between 2020 August 11 (MJD 59072) and October 16 (MJD 59138). The observations were performed with all three instruments onboard: the X-ray Telescope \citep[XRT;][0.2--10.0 keV]{burrows05}, the Ultraviolet/Optical Telescope \citep[UVOT;][170--600 nm]{roming05}, and the Burst Alert Telescope \citep[BAT;][15--150 keV]{barthelmy05}.

XRT observations were performed in photon counting mode, except for the first observation performed on 2020 October 6 for which observations in windowed timing (WT) mode were carried out \citep[for a description of XRT read-out modes, see][]{hill04}. The XRT spectra were generated with the {\em Swift}-XRT data product generator tool at the UK Swift Science Data Centre\footnote{http://www.swift.ac.uk/user\_objects} \citep[for details, see][]{evans09}. Spectra having count rates higher than 0.5 counts s$^{-1}$ may be affected by pile-up. To correct for this effect, the central region of the image has been excluded, and the source image has been extracted with an annular extraction region with an inner radius that depends on the level of pile-up \citep[see e.g.,][]{moretti05}.

The hard X-ray flux of this source is usually below the sensitivity of the BAT instrument for daily short exposures. Based on the {\em Swift}-BAT Hard X-ray Transient Monitor \citep{krimm13}\footnote{https://swift.gsfc.nasa.gov/results/transients/weak/QSOB2200p420}, considering daily light curves, only in three cases a detection with a significance $>$ 3 $\sigma$ has been observed in the studied period, on 2020 August 28 (MJD 59089), September 1 (MJD 59093), and September 13 (MJD 59105), with a corresponding count rate of (3.97 $\pm$ 1.11) $\times$10$^{-3}$, (4.54 $\pm$ 1.27) $\times$10$^{-3}$, and (3.35 $\pm$ 1.05) $\times$10$^{-3}$ counts cm$^{-2}$ s$^{-1}$, respectively. BL Lacertae is also included in the {\em Swift}-BAT 105-month hard X-ray catalogue \citep{oh18}, for observations carried out between 2004 December and 2013 August, with a photon index of $\Gamma$ = 1.76\,$^{+0.18}_{-0.17}$.

During the {\em Swift} pointings, the UVOT instrument observed the sources in its optical ($v$, $b$, and $u$) and UV ($w1$, $m2$, and $w2$) photometric bands \citep{poole08,breeveld10}. The UVOT data in all filters were analysed with the \texttt{uvotimsum}  and \texttt{uvotmaghist} tasks and the 20201215 CALDB-UVOTA release. Source counts were extracted from a circular region of 5 arcsec radius centred on the source, while background counts were derived from a circular region with a 20 arcsec radius in a nearby source-free region. All UVOT exposures were checked for possible small-scale sensitivity problems, which occur when the source falls on small detector regions where the sensitivity is lower\footnote{https://swift.gsfc.nasa.gov/analysis/uvot\_digest/sss\_check.html}. 

\subsection{{\em NICER} observations}\label{NICER_obs} 

The {\em NICER} \citep{gendreau12} on the {\em International Space Station} observed BL Lacertae for a Discretionary Data Time (DDT) request (PI: D'Ammando; ObsIds: 3201820101 and 3201820102) between 2020 October 11 17:35:04 UTC and October 12 23:13:04 UTC (MJD 59133.73268519--59134.96740741) for an effective time of 14.7 ks. The {\em NICER} observations were reduced using \texttt{NICERDAS} v7a and the calibration files available in the CALDB release 20200727. Good time intervals (GTI) were generated using \texttt{nimaketime} to select events that occurred when the particle background was low (KP\footnote{KP is the space weather index Kennziffer Planetary \citep{bartels39}. KP is derived from a worldwide network of magnetometers: https://www.swpc.noaa.gov/products/planetary-k-index} $<$ 5 and COR\_SAX\footnote{The COR\_SAX parameter estimates the magnetic cut-off rigidity.} $> 1.5$), and avoiding times of high optical loading. Moreover, the NICER focal plane modules 34, 14, 43, and 54 show episodes of increased detector noise. Data from these detectors have been excluded from the final events file. 

The background was estimated using the tool \texttt{nibackgen3C50}\_v6 \citep{remillard21}\footnote{https://heasarc.gsfc.nasa.gov/docs/nicer/tools/nicer\_bkg\_est\_tools.html}. Comparing the results to the background obtained using the tool \texttt{nicer\_bkg\_estimator}\_v6 (Gendreau et al., in preparation), we did no find significant differences. The event file filtered using the GTI is loaded into \texttt{XSELECT} to extract the source spectrum and a 240 s light curve. The source spectrum is rebinned with a minimum of 20 counts per energy bin with \texttt{grppha} to allow for $\chi^{2}$ spectrum fitting. 

\subsection{{\em NuSTAR} observations}\label{NUSTAR_obs}

The {\em NuSTAR} \citep[NuSTAR;][]{harrison13} observed BL Lacertae for a DDT proposal (PI: D'Ammando; ObsId: 90601630002) between 2020 October 11 14:06:09 UTC and October 12 07:16:09 UTC (MJD 59133.58760417--59134.30288194) with its two coaligned X-ray telescopes with corresponding focal planes, focal plane module A (FPMA) and B (FPMB), for 30.7 and 30.4 ks, respectively. The level 1 data products were processed with the {\em NuSTAR} Data Analysis Software (\texttt{nustardas}) package (v1.9.2). Cleaned event files (level 2 data products) were produced and calibrated using standard filtering criteria with the \texttt{nupipeline} software module, and the OPTIMIZED parameter for the exclusion of the South Atlantic Anomaly (SAA) passages. We used the calibration files available in the {\em NuSTAR} CALDB version 20210202. 

Spectra of the source were extracted for the whole observation from the cleaned event files using a circle of 30 pixel ($\sim$70 arcsec) radius, while the background was extracted from a nearby circular region of 30 pixel radius on the same chip of the source. The choice of the extraction region size optimizes the signal-to-noise, but alternative choices do not affect the results. The ancillary response files were generated with the \texttt{numkarf} task, applying corrections for the point spread function losses, exposure maps, and vignetting. The spectra were rebinned with a minimum of 20 counts per energy bin to allow for $\chi^{2}$ spectrum fitting. The net count rate for the entire observation is 0.193 $\pm$ 0.003 and 0.182 $\pm$ 0.003 counts s$^{-1}$ for FPMA and FPMB, respectively. The target is detected above the background in both the focal plan modules up to $\sim$75 keV. 

The observation is also divided by orbits, identified by the satellite's emergence from the SAA, as reported in the GTI file. The same procedure applied to the entire observation has been used for producing a spectrum for each satellite orbit.

\section{Analysis results}\label{results}

\begin{figure*}
\begin{center}
\rotatebox{0}{\resizebox{!}{82mm}{\includegraphics{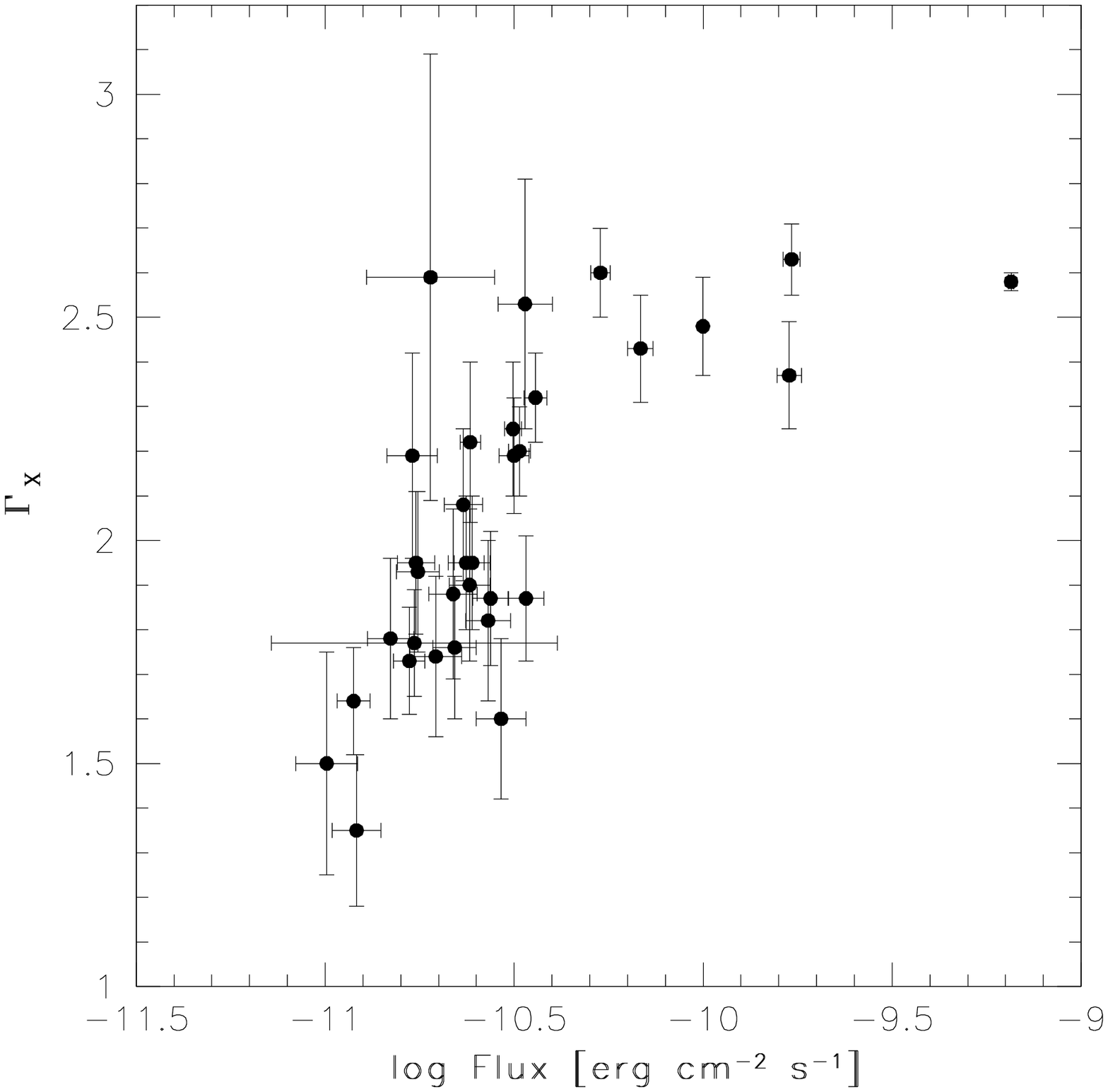}}}
\hspace{0.1cm}
\rotatebox{0}{\resizebox{!}{83mm}{\includegraphics{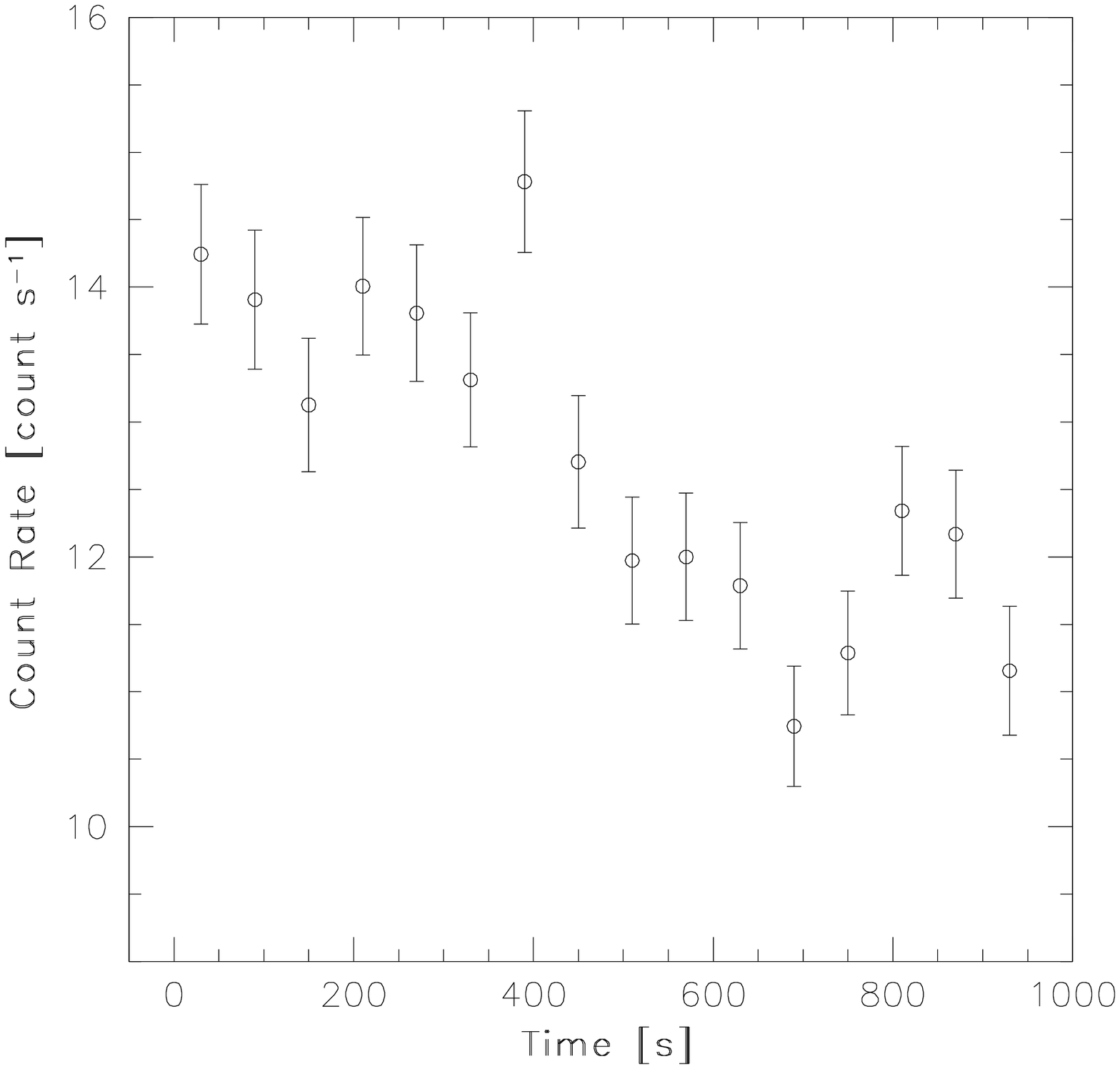}}}
\caption{{\it Left}: {\em Swift}-XRT photon index as a function of the 0.3--10 keV unabsorbed flux of BL Lacertae. {\it Right}: {\em Swift}-XRT light curve of BL Lacertae collected on 2020 October 6 is shown in terms of count rate using 60 s time bins.}
\label{XRTbis}
\end{center}
\end{figure*}

\subsection{{\em Swift}-XRT}\label{XRT_results}

The X-ray spectra collected by XRT in the 0.3--10 keV energy range are fitted by an absorbed power-law model using the photoelectric absorption model \texttt{tbabs} and set abundances according to \citet{wilms00} and \texttt{vern} cross-sections \citep{verner96}. The same abundances and cross-sections are also used for fitting the {\em NICER} and {\em NuSTAR} spectra later. The Galactic column density in the direction of the source has been initially fixed to $N_{H}$ = 2.70$\times$10$^{21}$ cm$^{-2}$, in agreement with the value used in \citet{madejski99} and \citet{weaver20} (but see Section \ref{summary} for a discussion about the total hydrogen column density towards BL Lacertae). We used the spectral redistribution matrices in the Calibration data base maintained by HEASARC. The X-ray spectral analysis was performed using the \texttt{XSPEC 12.9.1} software package \citep{arnaud96}. Data were grouped by single photons with \texttt{grppha} and the Cash statistics \citep{cash79} is used. The results of the fits are reported in Table~\ref{A1}. 

We plotted in Fig.~\ref{XRT_lc} the X-ray flux (top panel) and photon index (bottom panel) estimated in the 0.3--10 keV energy range with XRT observations. The X-ray flux corrected for Galactic extinction varies between 1.01$\times$10$^{-11}$ and 65.28$\times$10$^{-11}$ erg cm$^{-2}$ s$^{-1}$, with a median value of 2.42$\times$10$^{-11}$ erg cm$^{-2}$ s$^{-1}$. The corresponding variability amplitude, $V_{\mathrm{amp}}$, calculated as the ratio of maximum to minimum flux, is 64.6. As a comparison, the $V_{\mathrm{amp}}$ estimated between 2012 October 27 and 2013 November 5, at the time of the 2012 December flare, was 6.7 \citep{wehrle16}. A similar variability amplitude ($V_{\mathrm{amp}}$ $\sim$7) has been observed in X-rays considering the period 2008 August 4--2012 October 31 \citep{raiteri13}. 

In addition, we estimated the fractional variability parameter, $F_{\mathrm{var}}$, for taking into account also the uncertainties on the flux. We followed the prescription given by \citet{vaughan03}:

\begin{equation}
  F_{\mathrm{var}} = \sqrt{\frac{S^2 -
      <\sigma_{\mathrm{err}}^2>}{<F_{\rm\,X}>^2}} \nonumber
\end{equation}
\noindent where  $<F_{\rm\,X}>$ denotes the average X-ray flux, $S$ denotes the standard deviation of the $N$ flux measurements and \mbox{$<\sigma_{\mathrm{err}}^2>$} the mean squared error. 
The uncertainty of $F_{\mathrm{var}}$ is estimated following \citet{poutanen08}:

\begin{equation}
\Delta F_{\mathrm{var}} = \sqrt{F^{2}_{\mathrm{var}} +
  err(\sigma^{2}_{\mathrm{NXS}})} -F_{\mathrm{var}} \nonumber
\end{equation}

\noindent where $err(\sigma^{2}_{\mathrm{NXS}})$ is given by equation 11 in \citet{vaughan03}: 
\begin{equation}
err(\sigma^{2}_{\mathrm{NXS}}) = \sqrt{\left(\sqrt{\frac{2}{N}}\frac{<\sigma^{2}_\mathrm{err}>}{<F_{\rm\,X}>^{2}}\right)^{2}  +
  \left(\sqrt{\frac{<\sigma^{2}_\mathrm{err}>}{N}}\frac{2F_{\mathrm{var}}}{<F_{\rm\,X}>}\right)^{2}} \nonumber
\end{equation}

We obtained $F_{\mathrm{var}}$ = 0.44 $\pm$ 0.02, confirming the very high variability of the source in X-rays during 2020 August--October and showing that such variability is not dominated by the uncertainties on the flux. 

The photon index varies between 1.35 and 2.63, with a median value of  $\langle \Gamma_{\rm\,X} \rangle$ = 1.95 $\pm$ 0.12. A softer-when-brighter behaviour has been observed by XRT during 2020 August--October (see Fig.~\ref{XRTbis}, left-hand panel). To statistically investigate whether X-ray data follow this trend, a correlation study between flux and photon index is performed, using the Spearman rank test. We obtain a Spearman rank order correlation coefficient r$_{s}$ = 0.71, indicating a strong correlation. For N = 33 observations, there is a 0.1 per cent probability that the null hypothesis is correct (p = 0.001), corresponding to a 99.9 per cent statistical significance.

The flux observed on 2020 October 6 is the historical maximum reached by the source in X-rays\footnote{See also https://www.swift.psu.edu/monitoring/source.php?source\\=BLLacertae}. As a comparison, the maximum 0.3--10 keV flux observed during the 2012 flaring activity was 8.02$\times$10$^{-11}$\,erg cm$^{-2}$ s$^{-1}$ \citep{wehrle16}. That value is not corrected for Galactic absorption, so for a direct comparison the peak flux observed on 2020 October 6 not corrected for Galactic absorption is (36.89 $\pm$ 0.47) $\times$10$^{-11}$\,erg cm$^{-2}$ s$^{-1}$, a factor of 4.5 higher than the 2012 peak flux. 

Correcting for instrumental artefacts (i.e. hot pixels and bad columns on the CCD), pile-up, and after the background has been subtracted, we produced a light curve with time bins of 60 s for the XRT observation performed on October 6, i.e. at the peak of the activity (see Fig.~\ref{XRTbis}, right-hand panel). A significant change of the count rate ($>$ 3 $\sigma$)\footnote{\label{significance}The significance is calculated as $\Delta$CR/$\sqrt{(\sigma_{\rm\,CR1}^{2}+\sigma_{\rm\,CR2}^{2})}$, where $\Delta$CR = $|$count rate$_{\rm\,1}$--count rate$_{\rm\,2}$$|$, and $\sigma_{\rm\,CR1}$ and $\sigma_{\rm\,CR2}$ are the corresponding uncertainties.} has been observed in consecutive bins on a time-scale of 60 s. Following \citet{saito13}, to calculate the minimum doubling/halving time-scale between two consecutive points we used $\tau$ = $\Delta$t $\times$ ln2 / ln(CR(t2)/CR(t1)), where CR(t1) and CR(t2) are the count rate at time t1 and t2, respectively. We found $\tau$ = 274 s (256 s in the source rest frame)\footnote{The corresponding exponential growth time-scale is $\tau$/ln2 = 396 s.}. This value should be considered an upper limit limited by the number of photons collected in light curves produced with shorter time bins. In case of the XRT light curve produced with time bins of 15 s (see Fig.~\ref{A1bis}), the minimum doubling/halving time is $\tau$ = 50 s (47 s in the source rest frame). However, episodes of change of the count rate between consecutive bins are detected only at a significance level of 2 $<$ $\sigma$ $<$3.

On 2020 October 6, using an absorbed power law with $N_{H}$ fixed to 2.70$\times$10$^{21}$\,cm$^{-2}$, we obtained a photon index of $\Gamma$ = 2.58 $\pm$ 0.03 ($\chi^{2}$/dof = 404.34/254), while leaving the $N_{H}$ free to vary a photon index of $\Gamma$ = 2.77 $\pm$ 0.06 and $N_{H}$ = (3.8 $\pm$ 0.4) $\times$10$^{21}$\,cm$^{-2}$ are obtained with a significant improvement of the fit ($\chi^{2}$/dof = 264.46/253). Fixing the photon index to 2.58 and leaving the Galactic absorption free to vary, we obtained $N_{H}$ = (3.3 $\pm$ 0.1) $\times$10$^{21}$\,cm$^{-2}$ with an intermediate quality of fit with respect to the previous two fits ($\chi^{2}$/dof = 312.21/253).

We also summed the XRT observations carried out in two periods: (i) before the peak of the activity, i.e. 2020 September 6--25, for a total exposure of 6603 s; (ii) after the peak of the activity, i.e. 2020 October 11--16, for a total exposure of 8574 s. Fitting the spectra with an absorbed power law with $N_{H}$ fixed to 2.70$\times$10$^{21}$\,cm$^{-2}$, we obtained a photon index of $\Gamma$ = 2.08 $\pm$ 0.06 ($\chi^{2}$/dof = 101.40/90) and 1.71 $\pm$ 0.07 ($\chi^{2}$/dof = 73.32/72) for period (i) and (ii), respectively. Leaving the Galactic absorption value free to vary, we obtained $\Gamma$ = 1.94 $\pm$ 0.10 and $N_{H}$ = (2.1 $\pm$ 0.3) $\times$10$^{21}$\,cm$^{-2}$ ($\chi^{2}$/dof = 94.63/89), and $\Gamma$ = 1.65 $\pm$ 0.11 and $N_{H}$ = (2.4 $\pm$ 0.5) $\times$10$^{21}$\,cm$^{-2}$ ($\chi^{2}$/dof = 71.73/71) for period (i) and (ii), respectively. Fitting simultaneously the XRT spectra collected during the peak activity, periods (i) and (ii), fixing the photon index to 2.58, 2.08, and 1.71, respectively, and leaving the $N_{H}$ value free to vary, we obtained a value of (3.1 $\pm$ 0.1) $\times$10$^{21}$\,cm$^{-2}$.

\begin{figure}
\begin{center}
\includegraphics[width=0.5\textwidth]{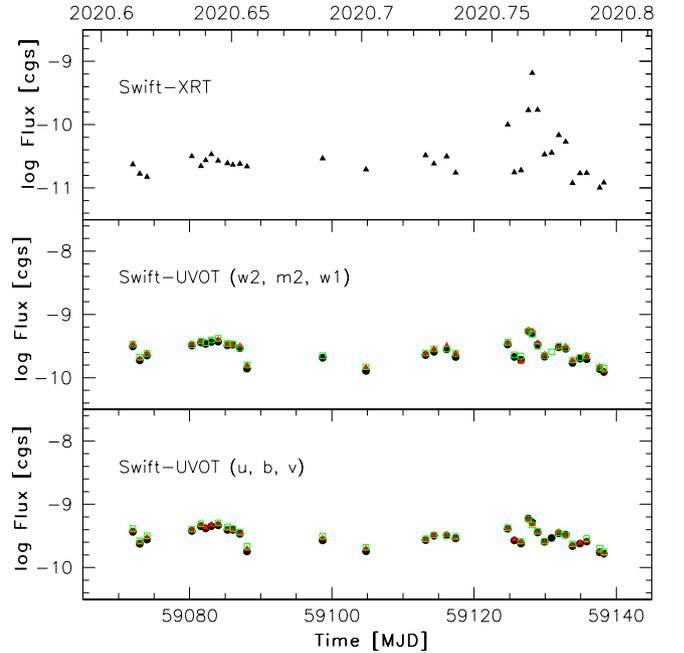}
\caption{Comparison of the light curve of BL Lacertae in X-ray (top panel; 0.3--10 keV), UV (middle panel; $w2$: black filled circles, $m2$: red filled triangles, $w1$: green open squares), and optical bands (bottom panel; $u$: black filled circles, $b$: red filled triangles, $v$: green open squares) collected by {\em Swift} during 2020 August 11--October 16. All values are corrected for Galactic extinction. The contribution of the host galaxy has been subtracted in the optical and UV bands. Errors are smaller than symbols; therefore, they are not shown in the plot.}
\label{Swift_lc}
\end{center}
\end{figure} 

\begin{figure}
\begin{center}
\includegraphics[width=0.5\textwidth]{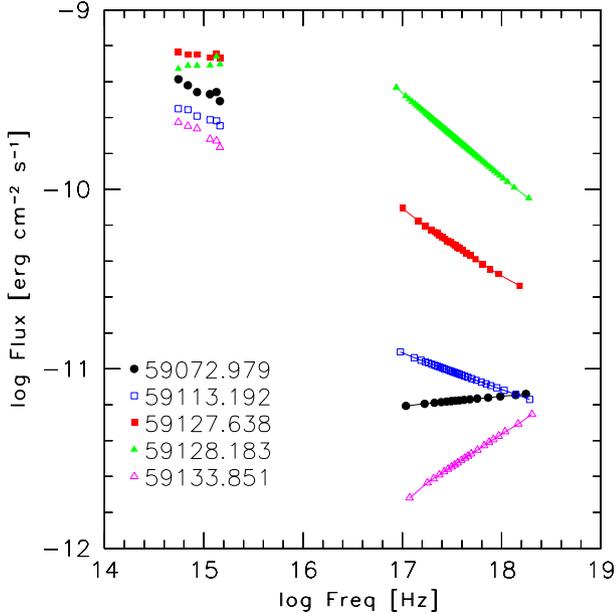}
\caption{SED of BL Lacertae from optical to X-rays collected by {\em Swift}-XRT and {\em Swift}-UVOT in five different epochs. The UVOT data are corrected for Galactic extinction and the contribution from the host galaxies has been subtracted. In X-rays the best-fitting model of the spectra collected has been shown. Different symbols and colours refer to different observational epochs.}
\label{Swift_SED}
\end{center}
\end{figure}

\subsection{{\em Swift}-UVOT}\label{UVOT_results}

Optical and UV observations obtained with the UVOT telescope simultaneously to XRT are useful to investigate variability properties, to build optical-to-X-ray SED at different epochs and to study the connection with the X-ray activity. The observed magnitudes of the source are reported in Table~\ref{UVOT}. Following \citet{raiteri13}, we assumed a flux density of 2.89, 1.30, 0.36, 0.026, 0.020, and 0.017 mJy for the host galaxy in the $v$, $b$, $u$, $w1$, $m2$, and $w2$ bands. By considering the source extraction radius used, the host galaxy contribution contaminating the UVOT photometry is about 50 per cent of the total galaxy flux, and it is removed from the magnitude values for calculating the flux densities. UVOT flux densities are also corrected for dust extinction using the E(B--V) value of 0.291 from \citet{schlafly11} and the extinction laws from \citet{cardelli89}. 

The variability amplitude estimated with the UVOT data is 3.32, 3.47, 3.60, 3.85, 4.39, 4.40 in the $v$, $b$, $u$, $w1$, $m2$, $w2$ bands, respectively, with an increasing variability going from the $v$ to the $uvw2$ band. This behaviour has been already observed in this source \citep[see e.g.][]{raiteri10,raiteri13}, confirming a dominant contribution of the beamed synchrotron emission from the jet with respect to the thermal emission from the accretion disc also in the UV band, differently from what is observed in some FSRQ \citep[e.g.,][]{raiteri12, orienti20}. 

The unabsorbed flux densities corrected for the host galaxy contribution are plotted in Fig.~\ref{Swift_lc} for a comparison with the unabsorbed X-ray fluxes observed during 2020 August--October. The source is less variable in optical and UV bands with respect to X-rays, showing a similar variability pattern in the three bands. It is interesting to notice that, although the flux density in optical and UV is high at the time of the peak of the X-ray activity (i.e., on August 6), the optical and UV activity peaked on August 5. This is in agreement with a significant shift of the synchrotron peak to higher frequencies, thus an increase of the contribution of the synchrotron component in the X-ray band passing from August 5 to 6, corresponding to a larger increase of the X-ray flux on August 6. Comparing the optical-to-X-ray SED of the source collected at different epochs (see Fig.~\ref{Swift_SED}), such a shift of the synchrotron peak is evident not only from August 5 to 6 but also between different activity states. With respect to the periods of relatively low activity, before (i.e. 2020 August 11; MJD 59072) and after (i.e. 2020 October 11; MJD 59133) the peak, at the time of the peak of the activity (i.e. 2020 October 5 and 6; MJD 59127 and MJD 59128) the optical--UV part of the SED is flatter, while the X-ray spectrum is softer ($\Gamma_{\rm\,X}$ $>$ 2).

\begin{figure}
\begin{center}
\includegraphics[width=0.5\textwidth]{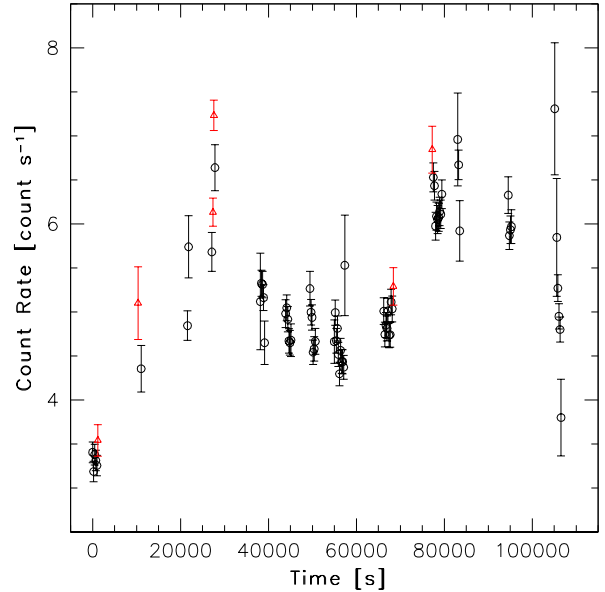}
\caption{{\em NICER} light curve of BL Lacertae shown in terms of count rate using 240 s time bins starting from T$_{0}$ = MJD 59133.58760417 (2020 October 11 17:35:04 UTC). Red triangles refer to periods of significant change of activity in consecutive bins.}
\label{NICER_lc}
\end{center}
\end{figure}

\subsection{{\em NICER}}

\begin{figure}
\begin{center}
\includegraphics[width=50mm, angle=270]{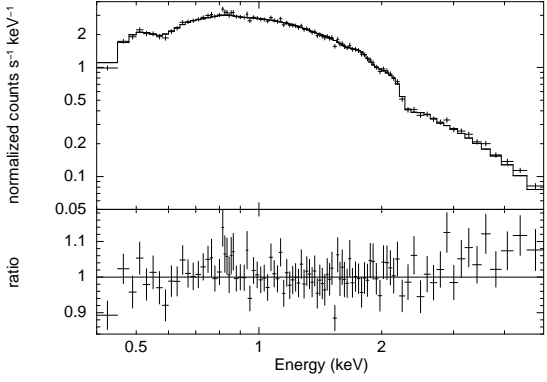}
\caption{{\em NICER} spectrum of BL Lacertae collected on 2020 October 11--12 in the 0.4--5 keV energy range fitted with a log-parabola model and the Galactic absorption fixed to 2.70$\times$10$^{21}$\,cm$^{-2}$.}
\label{NICER_logp}
\end{center}
\end{figure}

The source is detected above the background across the energy interval 0.4--5.0 keV, with an average count rate of 4.10 counts s$^{-1}$. We produced a light curve using 240 s time bins (Fig.~\ref{NICER_lc}). The 0.4--5 keV light curve shows variability by factors of up to 2 across the observation, with the count rate varying between 3.15 and 7.30 counts s$^{-1}$ and episodes of variability on time-scales of a few thousand seconds. In particular, there are three episodes of significant change of activity ($>$ 3 $\sigma$; see footnote~\ref{significance}) in consecutive bins of the light curve, with changes from 3.54 $\pm$ 0.18 to 5.10 $\pm$ 0.41 counts s$^{-1}$ in 9120 s, from 6.13 $\pm$ 0.16 to 7.23 $\pm$ 0.17 counts s$^{-1}$ in 240 s, and from 5.29 $\pm$ 0.21 to 6.83 $\pm$ 0.21 counts s$^{-1}$ in 8840 s. It is particularly remarkable the increase of count rate observed in only 240 s considering that, differently from the other two cases, there are no observational gaps between the values estimated in consecutive bins. Assuming a redshift of $z$ = 0.069, the observed variability time-scale corresponds to 8531, 225, and 8269 s, respectively, in the source rest frame. The minimum doubling/halving time-scale for the {\em NICER} observations is $\tau$ =  1008 s (943 s in the source rest frame).

\begin{figure*}
\begin{center}
\rotatebox{0}{\resizebox{!}{72mm}{\includegraphics{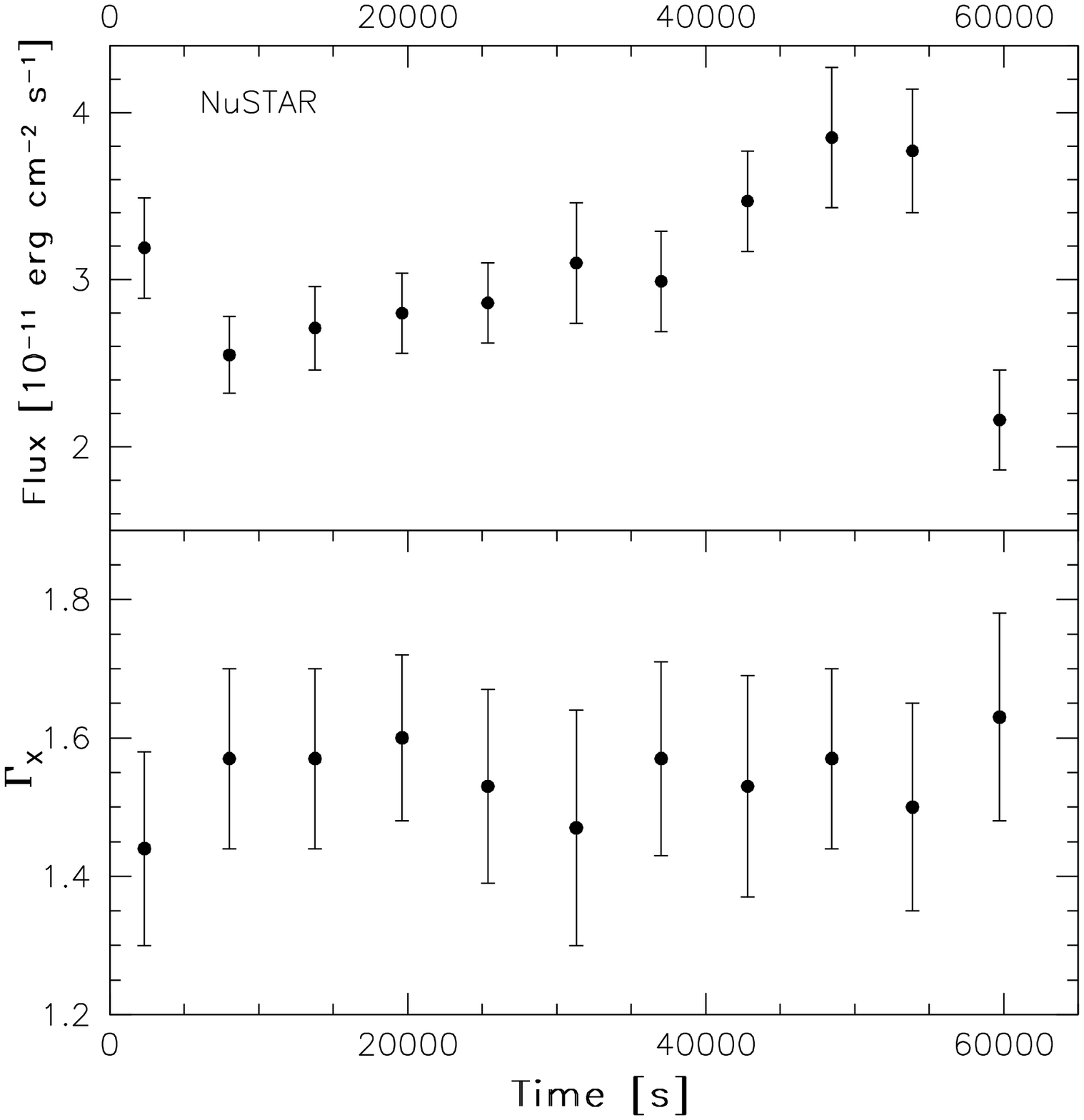}}}
\hspace{0.1cm}
\rotatebox{0}{\resizebox{!}{63mm}{\includegraphics{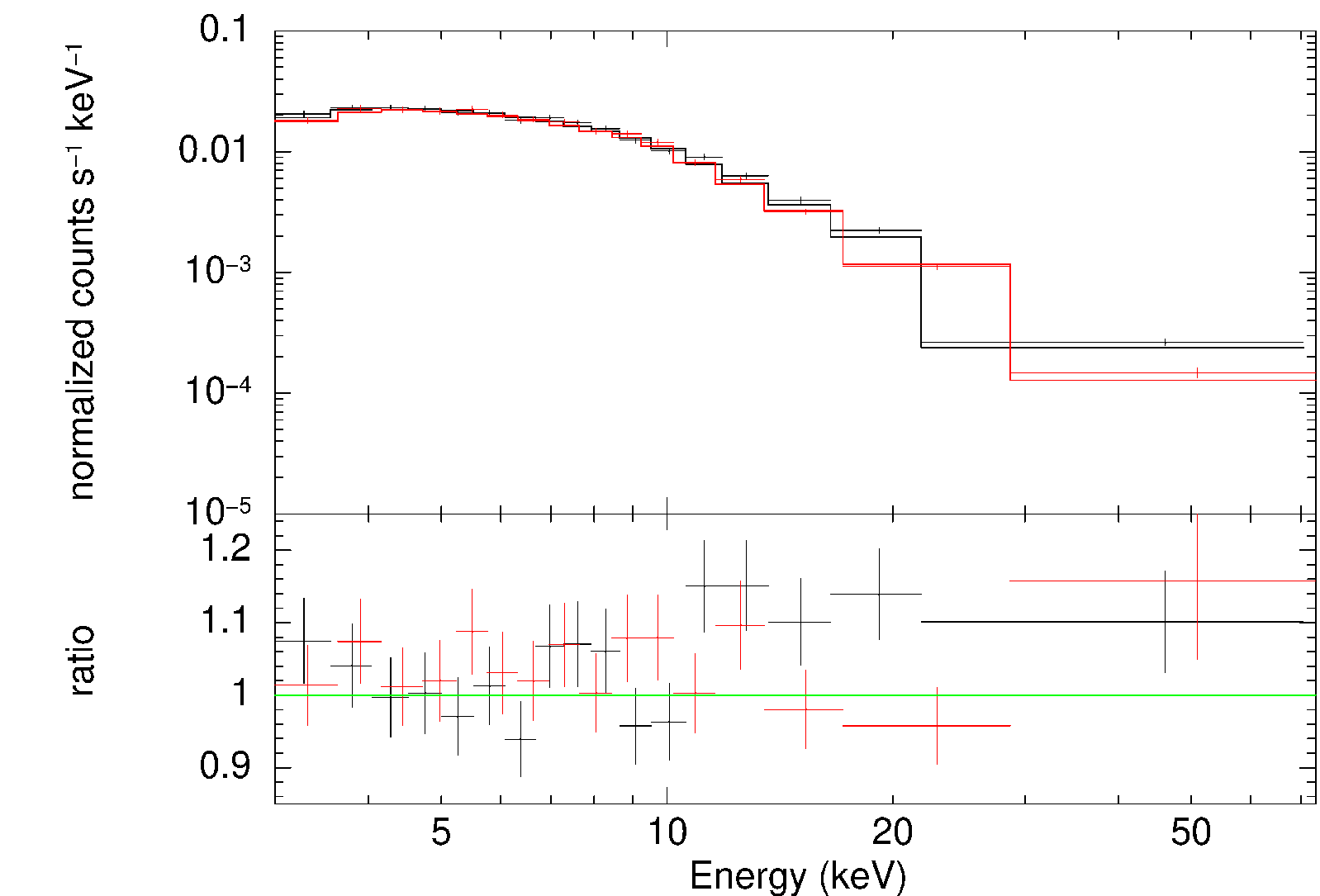}}}
\caption{{\it Left-hand panel}: {\em NuSTAR} light curve of BL Lacertae orbit-by-orbit as seen by the FPMA module with flux ({\it upper panel}) and photon index ({\it bottom panel}) in the 3.0--79 keV energy range. {\it Right-hand panel}: {\em NuSTAR} spectrum collected on 2020 October 11--12 in the 3--79 keV energy range fitted with a power-law model and the Galactic absorption fixed to 2.70$\times$10$^{21}$\,cm$^{-2}$.}
\label{NuSTAR_orbit_lc}
\end{center}
\end{figure*}

The {\em NICER} spectrum can be fitted in the 0.4--5.0 keV energy range with an absorbed power law with a photon index of $\Gamma$ = 2.06 $\pm$ 0.01 and the Galactic absorption corresponding to a hydrogen column density fixed to $N_{H}$ = 2.70$\times$10$^{21}$\,cm$^{-2}$ ($\chi^{2}$/dof = 527.11/457). A better fit can be obtained by using an absorbed log-parabola model with a slope $\alpha$ = 2.10 $\pm$ 0.02 and a negative curvature parameter of $\beta$ = $-$0.19 $\pm$ 0.05, suggesting a concave X-ray spectrum ($\chi^{2}$/dof = 490.80/456; see Fig.~\ref{NICER_logp}). Applying an F-test to compare the power-law and log-parabola models, we obtained a probability that the null hypothesis is true of 1.19$^{-8}$. The corresponding flux (corrected for Galactic absorption) in the 0.3--10 keV/0.4--5.0 keV energy range is (1.62 $\pm$ 0.01) $\times$10$^{-11}$ / (1.13 $\pm$ 0.01) $\times$10$^{-11}$\,erg cm$^{-2}$ s$^{-1}$. Compared to the XRT observations, the {\em NICER} observations have been carried out during an intermediate flux level (see Fig.~\ref{XRT_lc}, upper panel). Leaving the Galactic absorption value free to vary, in case of a log-parabola model, we found a slight improvement of the fit ($\chi^{2}$/dof = 481.13/455) with larger uncertainties on the spectral parameters, $\alpha$ = 2.32 $\pm$ 0.12 and $\beta$ = $-$0.48 $\pm$ 0.16, and $N_{H}$ = (3.1 $\pm$ 0.3)$\times$10$^{21}$\,cm$^{-2}$. 

\subsection{{\em NuSTAR}}

The {\em NuSTAR} observations extended over 11 satellite orbits. First, we investigated the behaviour orbit-by-orbit in the {\em NuSTAR} bandpass. A similar variability amplitude and behaviour have been observed by considering the count rate orbit-by-orbit in the 3.0--79, 3.0--10.0, and 10--79 keV energy range (see Fig.~\ref{B1}). For clarity, only the count rate from FPMA is shown. Data from FPMB showed a similar situation.

By fitting the spectra obtained for each orbit with a simple power law and $N_{H}$ fixed to 2.70$\times$10$^{21}$\,cm$^{-2}$, the photon index ranges between 1.44 and 1.63, with an average value of 1.54 and no significant variability observed within the uncertainties. The corresponding 3--79 keV fluxes range between 2.55$\times$10$^{-11}$ and 3.85$\times$10$^{-11}$\,erg cm$^{-2}$ s$^{-1}$, with an average value of 3.04$\times$10$^{-11}$\,erg cm$^{-2}$ s$^{-1}$ and an increase up to 50 per cent of the flux over the observation (see Table~\ref{NuSTAR_orbit} and Fig.~\ref{NuSTAR_orbit_lc}, left-hand panel). A significant ($>$3 $\sigma$) change of flux has been observed between the last two orbits, in particular on a time-scale of 5844 s (5468 s in the source rest frame). The minimum doubling/halving time-scale considering the orbit-by orbit {\em NuSTAR} observations is $\tau$ = 17650 s (16510 s in the source rest frame).

Previous {\em NuSTAR} observations of the source have been carried out in 2012 and 2019. During 2012 December 11--12, the orbit-by-orbit photon index varied between 1.81 and 1.93 with an average value of 1.86, and flux changes of up to 30 per cent \citep{wehrle16}. The 3--7 keV flux varied between 1.1 and 1.3$\times$10$^{-11}$\,erg cm$^{-2}$ s$^{-1}$. As a comparison, in our observation, the 3--7 keV flux varies between 3.6 and 5.5$\times$10$^{-12}$\,erg cm$^{-2}$ s$^{-1}$, a factor of 2--3 lower than the 2012 fluxes. On the contrary, during 2019 September 14--19, the 3--79 flux of the source estimated orbit-by-orbit has been lower than that in 2020 with values between 0.86 and 1.97$\times$10$^{-11}$\,erg cm$^{-2}$ s$^{-1}$, and the photon index ranged between 1.56 and 2.15 with an average value of 1.87 \citep{weaver20}. Therefore, compared to previous {\em NuSTAR} observations, the 2020 observations showed an intermediate hard X-ray flux between 2019 and 2012 observations and a harder photon index with respect to both of them.     

We fitted the overall {\em NuSTAR} spectrum by using a simple power law and a log-parabola model, assuming $N_{H}$ = 2.70$\times$10$^{21}$\,cm$^{-2}$. We also included a cross-normalization factor for FPMB with respect to FPMA that resulted in all cases $<$ 5 per cent, consistent with the expectations from calibration observations \citep{madsen15}. The simple power-law model gives acceptable results with a $\chi^{2}$/dof of 452.85/458 and a photon index of $\Gamma$ = 1.59 $\pm$ 0.03 (see Fig.~\ref{NuSTAR_orbit_lc}, right-hand panel). The corresponding 3--79 keV flux is (2.97\,$^{+0.05}_{-0.06}$)$\times$10$^{-11}$\,erg cm$^{-2}$ s$^{-1}$. A comparable quality of fit has been obtained by using a log-parabola model ($\chi^{2}$/dof = 452.84/457) with a slope $\alpha$ = 1.59 $\pm$ 0.04 and a negligible curvature parameter of $\beta$ = 0.003 $\pm$ 0.010. Fitting the spectrum with a broken power-law model does not improve the fit in this case either. No obvious spectral break is determined within the {\em NuSTAR} bandpass. 
 
\begin{table*}
\caption{Log and fitting results of {\em NuSTAR} orbit-by-orbit observations of BL Lacertae using a power-law model with $N_{\rm H}$ fixed to 2.70$\times$10$^{21}$\,cm$^{-2}$. The fit refers to FPMA only.} 
\label{NuSTAR_orbit}
\begin{center}
\begin{tabular}{ccccc}
\hline
\multicolumn{1}{c}{\textbf{Orbit}} &
\multicolumn{1}{c}{\textbf{Net exposure time}} &
\multicolumn{1}{c}{\textbf{Count rate}} &
\multicolumn{1}{c}{\textbf{Photon index}}      &
\multicolumn{1}{c}{\textbf{Flux$_{\rm\,3-79\,keV}$}}  \\
\multicolumn{1}{c}{} &
\multicolumn{1}{c}{(s)} &
\multicolumn{1}{c}{(counts s$^{-1}$)} &
\multicolumn{1}{c}{($\Gamma_{\rm\,X}$)}  &
\multicolumn{1}{c}{(10$^{-11}$ erg cm$^{-2}$ s$^{-1}$)}  \\
\hline
1  &  3196 & 0.169 $\pm$ 0.007 &  1.44 $\pm$ 0.14 & 3.19 $\pm$ 0.30 \\  
2  &  3205 & 0.159 $\pm$ 0.007 &  1.57 $\pm$ 0.13 & 2.55 $\pm$ 0.23 \\  
3  &  3168 & 0.183 $\pm$ 0.008 &  1.57 $\pm$ 0.13 & 2.71 $\pm$ 0.25 \\   
4  &  3193 & 0.183 $\pm$ 0.008 &  1.60 $\pm$ 0.12 & 2.80 $\pm$ 0.24 \\  
5  &  3151 & 0.180 $\pm$ 0.008 &  1.53 $\pm$ 0.14 & 2.86 $\pm$ 0.24 \\  
6  &  2770 & 0.183 $\pm$ 0.008 &  1.47 $\pm$ 0.17 & 3.10 $\pm$ 0.36 \\ 
7  &  2558 & 0.186 $\pm$ 0.009 &  1.57 $\pm$ 0.14 & 2.99 $\pm$ 0.30 \\ 
8  &  2422 & 0.214 $\pm$ 0.010 &  1.53 $\pm$ 0.16 & 3.47 $\pm$ 0.30 \\ 
9  &  2298 & 0.248 $\pm$ 0.011 &  1.57 $\pm$ 0.13 & 3.85 $\pm$ 0.42 \\
10 &  2235 & 0.227 $\pm$ 0.010 &  1.50 $\pm$ 0.15 & 3.77 $\pm$ 0.37 \\
11 &  2521 & 0.221 $\pm$ 0.010 &  1.63 $\pm$ 0.15 & 2.16 $\pm$ 0.30 \\
\hline
\end{tabular}                   
\end{center}
\end{table*}

\section{Joint {\em NICER} and {\em NuSTAR} fit}\label{joint}

\begin{figure*}
\begin{center}
\includegraphics[width=0.6\textwidth, angle=270]{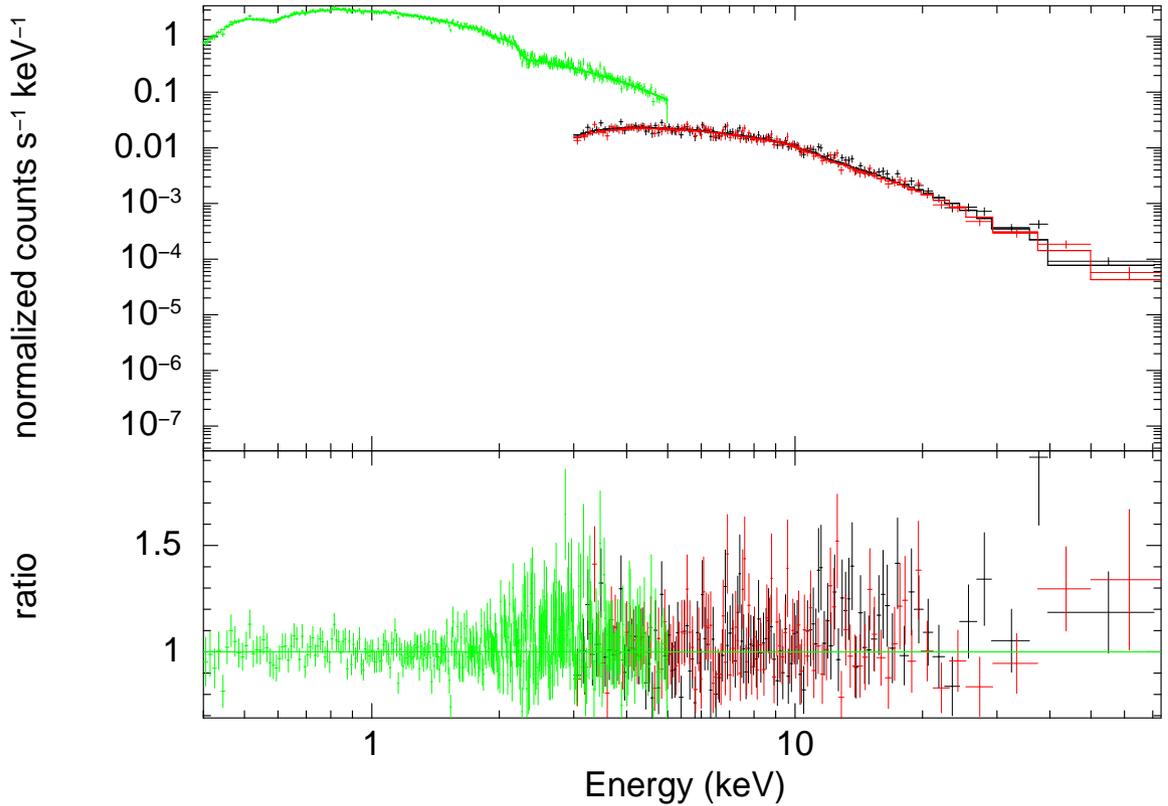}
\caption{{\em NICER} and {\em NuSTAR} spectra ({\it upper panel}) and residuals ({\it bottom panel}) of BL Lacertae collected on 2020 October 11--12 fitted in the 0.4--79~keV energy range with a broken power-law model with $N_{\rm H, tot}$ fixed to 2.70$\times$10$^{21}$\,cm$^{-2}$. Black, red, and green points represent {\em NuSTAR} FPMA, {\em NuSTAR} FPMB, and {\em NICER} data, respectively.}
\label{joint_fig}                                                                                                                                                                                     
\end{center}
\end{figure*}

\begin{table*}
\caption{\small{Summary of fits to the 0.4--79~keV {\em NICER} and {\em NuSTAR} joint spectrum of BL Lacertae. Asterisk refers to a fixed parameter.}}\label{jointfit}
\begin{center} 
\begin{tabular}{lll}
\hline \\ [-4pt]
Model & Parameter & Value \\[4pt]  
\hline\\[-6pt]
Power-law & $\Gamma$                                   &  1.98 $\pm$ 0.01 \\
          & $N_{\rm H, tot}$ (cm$^{-2}$) $^{*}$                    &  2.70$\times$10$^{21}$\,* \\ 
          & Cross-normalization FPMA/FPMB  &  1.06 $\pm$ 0.03 \\  
          & Cross-normalization FPMA/NICER &  0.76 $\pm$ 0.02 \\
          & $\chi^{2}$/dof                             & 1453.19/916      \\
\hline 
Log-parabola & $\alpha$ & $2.11 \pm 0.02$     \\
             & $\beta$  & $-$0.25 $\pm$ 0.02  \\
             & $N_{\rm H, tot}$ (cm$^{-2}$) $^{*}$                    &  2.70$\times$10$^{21}$\,* \\ 
             & Cross-normalization FPMA/FPMB  &  1.05 $\pm$ 0.03 \\    
             & Cross-normalization FPMA/NICER &  0.89 $\pm$ 0.03 \\
             & $\chi^{2}$/dof                             &   972.44/915 \\
\hline 
Two power-laws & $\Gamma_1$ & $2.09 \pm 0.02$     \\
               & $\Gamma_2$  & $0.87 \pm 0.06$  \\
               & $N_{\rm H, tot}$ (cm$^{-2}$) $^{*}$                    &  2.70$\times$10$^{21}$\,* \\ 
               & Cross-normalization FPMA/FPMB  &  1.09 $\pm$ 0.05 \\    
               & Cross-normalization FPMA/NICER &  0.84 $\pm$ 0.04 \\
               & $\chi^{2}$/dof                             &  1011.70/914 \\

\hline
Broken power-law & $\Gamma_1$ & 2.10 $\pm$ 0.02 \\
      & $E_{\rm\,break}$ (keV)     & 2.67\,$^{+0.29}_{-0.17}$ \\
      & $\Gamma_2$                 & 1.60\,$^{+0.03}_{-0.03}$ \\
      & $N_{\rm H, tot}$ (cm$^{-2}$) $^{*}$  &  2.70$\times$10$^{21}$\,* \\ 
      & Cross-normalization FPMA/FPMB  &  1.05 $\pm$ 0.03 \\
      & Cross-normalization FPMA/NICER &  0.97 $\pm$ 0.04 \\
      & $\chi^{2}$/dof                             &   954.69/914     \\
\hline
Broken power-law & $\Gamma_1$      & 2.04\,$^{+0.07}_{-0.04}$ \\
      & $E_{\rm\,break}$ (keV)     & 3.01\,$^{+0.41}_{-0.45}$ \\
      & $\Gamma_2$                 & 1.60\,$^{+0.03}_{-0.03}$ \\
      & $N_{\rm H, tot}$ (cm$^{-2}$)    &  2.59\,$^{+0.14}_{-0.09}$$\times$10$^{21}$ \\ 
      & Cross-normalization FPMA/FPMB  &  1.05 $\pm$ 0.03 \\
      & Cross-normalization FPMA/NICER &  0.96 $\pm$ 0.04 \\
      & $\chi^{2}$/dof                             &   953.17/913     \\
\hline      
\end{tabular}
\end{center}
\end{table*}

To investigate in detail the broad band X-ray spectrum of BL Lacertae, a joint fit to the lower energy {\em NICER} and higher energy {\em NuSTAR} data is performed. The combination of {\em NICER} and {\em NuSTAR} observations is important for characterizing the broad band spectrum and estimating the value of total hydrogen column density towards the source, not well determined so far. The {\em NICER} spectrum has a much better statistics than the {\em Swift}-XRT spectra collected on October 11 and 12 (74319 counts vs. 538 counts and 157 counts); for this reason, the XRT data are not used. The Galactic column density has been initially set to 2.70$\times$10$^{21}$\,cm$^{-2}$. A cross-normalization factor between the {\em NICER} and {\em NuSTAR} instruments has been added for taking into account differences in the absolute flux calibration and the slight offset of the observing times. A comparable median count rate is observed in the NICER light curve for the period simultaneous to the {\em NuSTAR} observation (4.94 counts s$^{-1}$) and just after that (5.01 counts s$^{-1}$; see Fig.~\ref{B2}), justifying the use of the entire {\em NICER} observation for the fitting. A simple power-law, a log-parabola, two power-laws, and a broken power-law model has been applied for the joint fitting of the {\em NICER} and {\em NuSTAR} data. The fitting results are summarized in Table~\ref{jointfit}.

The joint spectrum cannot be well represented from 0.4 to 79 keV by a simple power law  with a photon index $\Gamma$ = 1.98 $\pm$ 0.01 ($\chi^{2}$/dof = 1453.19/916). A log-parabola model has been also tested, with the pivot energy fixed at 1 keV, and the fit significantly improves ($\chi^{2}$/dof = 972.44/915) with a slope $\alpha$ = 2.11 $\pm$ 0.02 and a negative curvature $\beta$ = $-$0.25 $\pm$ 0.02, suggesting a concave spectrum as in the case of the fit of the {\em NICER} spectrum alone. By applying an F-test, we obtain a probability of 7.0$\times$10$^{-82}$ that the null hypothesis is true. Using two power-law models, the quality of the fit is worse than the case of the log-parabola model ($\chi^{2}$/dof = 1011.70/914), with the photon index of the second power law very hard ($\Gamma_2$ = 0.87).

A further improvement of the fit is obtained by using a broken power-law model ($\chi^{2}$/dof = 954.69/914). By applying an F-test, the improvement of the fit using a broken power law with respect to a power law is even more significant than a log-parabola model, with a probability that the null hypothesis is true of 4.1$\times$10$^{-84}$. We obtained photon indices of $\Gamma_{1}$ = 2.10 $\pm$ 0.02 and $\Gamma_{2}$ = 1.60 $\pm$ 0.03 below and above an energy break $E_{\rm\,break}$ = 2.67\,$^{+0.29}_{-0.17}$, respectively. The exposure of {\em NICER} and {\em NuSTAR} significantly overlapped, in agreement also with the cross-normalization obtained (0.97 $\pm$ 0.04 between FPMA and {\em NICER}). Fig.~\ref{joint_fig} shows the joint fit to the {\em NICER} and {\em NuSTAR} data for the simultaneous observations performed on 2020 October 11--12. 

Previous {\em NuSTAR} observations of the source in 2012 and 2019 have been carried out contemporaneously to {\em Swift}-XRT observations, and the $N_{\rm H, tot}$ value has not been left free to vary during the fitting procedure, probably due to the small number of counts collected by XRT in the low-energy part of the X-ray spectrum. Thanks to the {\em NICER} observations, we are able to test the total hydrogen column density for BL Lacertae combining the {\em NICER} and {\em NuSTAR} observations over a broad band energy range. Leaving the Galactic total absorption free to vary, the fit marginally improved ($\chi^{2}$/dof = 953.17/913), with $N_{\rm H, tot}$ = (2.59\,$^{+0.14}_{-0.09}$)$\times$10$^{21}$\,cm$^{-2}$ (see Table~\ref{jointfit}). Fixing the spectral parameters of the broken power-law model ($\Gamma_{1}$, $\Gamma_{2}$, $E_{\rm\,break}$) and leaving the Galactic total absorption free to vary, we obtained $N_{\rm H, tot}$ = (2.69\,$^{+0.04}_{-0.04}$)$\times$10$^{21}$\,cm$^{-2}$ with a comparable quality of the fit ($\chi^{2}$/dof = 954.54/916). 

\section{Discussion and conclusions}\label{summary}

Considering the large flaring activity of BL Lacertae observed from optical to $\gamma$-rays during 2020 August--October, that is an important possibility to study in detail the spectrum of the source at different frequencies. In this paper, we have analysed the X-ray data collected by {\em Swift}-XRT, {\em NICER}, and {\em NuSTAR}. These data allowed us to investigate the spectral and flux variability on different time-scales and to characterize in detail the broad band energy range of the source from 0.4 to 79 keV. We have reported the results of the first {\em NICER} observation of the source and, more generally, the first {\em NICER} and {\em NuSTAR} simultaneous follow-up observations of a $\gamma$-ray flaring blazar.

{\em Swift}-XRT monitored BL Lacertae between 2020 August 11 and October 16, observing on October 6 the historical maximum activity from this source in X-rays, with an unabsorbed flux of 6.53$\times$10$^{-10}$\,erg cm$^{-2}$ s$^{-1}$ and a corresponding luminosity of (7.86 $\pm$ 0.08) $\times$10$^{45}$\,erg s$^{-1}$ in the 0.3--10 keV energy range. The $V_{\rm\,amp}$ estimated over the period 2020 August--October is 64.6, a factor of 10 higher than the value observed at the time of the 2012 flaring activity, clearly indicating how exceptional is the X-ray flaring activity observed in 2020 October. 

\noindent Moreover, considering the {\em Swift}-XRT monitoring of other 10 bright $\gamma$-ray BL Lacs during 2004 December--2012 August, only Mrk 421 has shown a variability amplitude \citep[$V_{\mathrm{amp}}$ = 81.5;][]{stroh13} larger than the value estimated in 2020 for BL Lacertae. However, Mrk 421 is an HBL with a peak of the synchrotron emission usually in the soft X-ray energy range; therefore, even assuming the same level of activity, a larger variability amplitude is expected in that energy range for Mrk 421 with respect to IBL/LBL sources, like BL Lacertae. A similar variability pattern has been observed with XRT and UVOT in optical, UV, and X-rays with less variability in optical and UV bands with respect to the X-ray one.

During the XRT monitoring of BL Lacertae, we observed a softer-when-brighter behaviour in X-rays, with the photon index ranges between 1.35 and 2.63. This behavior can be related to an increasing importance of the synchrotron emission in the X-ray part of the spectrum covered by XRT during bright states, likely due to a shift of the synchrotron and IC peaks to higher frequencies (see Fig.~\ref{Swift_SED}). In this context, we noticed that an X-ray photon index higher than 2.2 (see Table A1) has been estimated in the two XRT observations close to the VHE detection of the source reported by the MAGIC telescopes on 2020 August 19 \citep{blanch20a} and September 19 \citep[][; see the SED collected on 2020 September 21, MJD 59113, in Fig.~\ref{Swift_SED}]{blanch20b}, in agreement with a shift of both the SED peaks to higher frequencies in these periods. A similar softer-when-brighter behaviour has been reported in \citet{weaver20}, when 40 XRT observations of BL Lacertae have been carried out between 2019 September 14 and 19. In that period, the fluxes were lower than the values observed in 2020 August--October [i.e. (3.1$\times$10$^{-12}$) -- (1.8$\times$10$^{-11}$)\,erg cm$^{-2}$ s$^{-1}$] with the photon index ranges from 1.79 to 2.72. A strong softer-when-brighter spectral variability pattern has been observed in OJ 287 \citep{komossa17,komossa21}, a BL Lac object classified as LSP or ISP, similarly to BL Lacertae.
On the contrary, a harder-when-brighter behaviour for BL Lacertae has been reported by \citet{wehrle16} during the 2012 flaring activity. This can be an indication that different emission mechanisms and/or changes of distinct jet parameters are at work in the source during different flaring activities. The combination of the data analysed here with other multi-wavelength data collected during this period, in particular by {\em Fermi}-LAT and MAGIC, will be important to study in detail this behaviour. 

Rapid variability in soft X-rays has been observed with both {\em Swift}-XRT and {\em NICER} in 2020 October. In particular, at the peak of the activity (October 6) a minimum variability time-scale of 60 s and a doubling time-scale of 274 s (256 s in the source rest frame) have been observed by XRT. Based on causality argument, it is possible to constrain the intrinsic size of the emitting region to be R $<$ $c$ $\delta$ $\tau$ /(1+$z$), where $\tau$ is the minimum doubling time-scale observed and $\delta$ is the Doppler factor. Assuming a typical Doppler factor of $\delta$ = 15 \citep[see e.g.][]{raiteri13}, we obtain R $<$ 1.1$\times$10$^{14}$ cm. {\em NICER} observations performed during an intermediate X-ray flux level, a few days after the peak of activity, observed a minimum variability time-scale of 240 s and a doubling time-scale of 1008 s (943 s in the source rest frame). This corresponds to an emitting region R $<$ 4.0$\times$10$^{14}$ cm. 

\noindent The light-crossing time for a Kerr black hole is t$_{\rm lc}$ = 2G\,M$_{\rm BH}$\,/$c^{3}$ = 2$\times$10$^{3}$\,(M$_{\rm BH}$/10$^{8}$\,M$_{\odot}$)\,s. Assuming a black hole mass of M$_{\rm BH}$ = 1.7$\times$10$^{8}$\,M$_{\odot}$ for BL Lacertae \citep{woo02}, we have t$_{\rm lc}$ = 3400 s. After considering the relativistic Doppler effect, we have t$_{\rm lc}$/$\delta$, and thus for $\delta$ = 15 we obtain 227 s, comparable to the minimum doubling time-scale observed by {\em Swift}-XRT on October 6 for the light curve produced with time bins of 60 s. However, a hint of shorter doubling time-scale (i.e. 50 s) has been observed inspecting the light curve produced with 15 s time bins. A Doppler factor significantly higher, i.e. 70, is needed to reconcile the light-crossing times of the black hole with such a short variability time-scale. We are in a similar situation if we assume a black hole mass of M$_{\rm BH}$ $\sim$ 5$\times$10$^{8}$\,M$_{\odot}$, as obtained by \citet{ghisellini10} and \citet{falomo03}. The rapid increase of the X-ray activity observed by XRT can indicate that in case of short time-scale variability the size of the black hole is not a hard lower limit on the physical size of the emitting region. In fact, the time-scales observed by XRT suggest that the emission is not produced by the entire jet but by compact regions within the jet. Different scenarios can explain such a small compact region: `jet-in-a-jet models' including ultra-relativistic outﬂow of material from magnetic reconnection sites \citep[e.g.,][]{giannios09} or relativistic turbulence in the jet \citep[e.g.,][]{narayan12}, or turbulent extreme multi-zone models \citep[e.g.,][]{marscher14}. In 1999 June, {\em BeppoSAX} observed that the soft X-ray flux of the source doubled on a time-scale of $\sim$20 m \citep{ravasio02}. Therefore, the minimum variability time-scales observed in 2020 October by {\em Swift}-XRT and {\em NICER} in the soft X-ray band are the most extreme observed from this source so far. More generally, fast X-ray variability has not been revealed in a large number of blazars \citep[e.g.,][]{pryal15}. This can be related to the fact that follow-up X-ray observations with different satellites are usually carried out from 1 d to a few days after the peak of the $\gamma$-ray flaring activity. In this way, we are usually observing the source in a high state but not at the peak of the activity when the variability would be extreme. 

At hard X-rays, {\em NuSTAR} observed a significant variability between two satellite's orbits ($\sim$5.8 ks), with a doubling time-scale of 17650 s (16510 s in the source rest frame). Previous {\em NuSTAR} observations in 2019 have shown a variability time-scale significantly longer of 14.5 h \citep{weaver20}.

The joint {\em NICER} and {\em NuSTAR} spectrum is well described by a broken power-law model with photon indices $\Gamma_{1}$ = 2.10 $\pm$ 0.02 and $\Gamma_{2}$ = 1.60 $\pm$ 0.03 below and above an energy break $E_{\rm\,break}$ = 2.67\,$^{+0.29}_{-0.17}$, respectively. There is a significant difference of the photon index estimated in the {\em NICER} and {\em NuSTAR} observations alone, in agreement with the results obtained for the joint spectrum applying a broken power-law model. This should be related to the presence of the high-frequency end of the synchrotron emission below a few keV, and the IC component dominating the emission at higher energies. The contemporaneous {\em NuSTAR} and {\em Swift}-XRT spectra of BL Lacertae collected on 2012 December 11--12, during another flaring activity, are well described by a broken power-law model with a photon index of 3.3$^{+1.3}_{-0.7}$ and 1.88 $\pm$ 0.01 below and above an energy break of 1.0 $\pm$ 0.2 keV \citep{wehrle16}. On the contrary, during the 2019 low-activity state no statistically significant improvement has been obtained for a broken power-law model over a single power-law model \citep{weaver20}. 

\noindent This source has been observed with several X-ray satellites in the past. Previous observations with ASCA \citep{sambruna99} and {\em BeppoSAX} \citep{ravasio03} have shown that a broken power law is statistically preferred over a single power-law model \citep{sambruna99}. In case of three {\em XMM-Newton} observations carried out in 2007--2008, a double power-law model represents better the 0.3--10 keV spectra with respect to a single power-law model \citep{raiteri09}. In that case, the photon indices of the two power-laws are $\Gamma_1$ = 2.48--2.58 and $\Gamma_2$ = 1.51--1.72, in agreement with the two photon indices obtained below and above the energy break of the joint fitting of the {\em NICER} and {\em NuSTAR} spectra with a broken power-law model presented here. 

The shape of the X-ray spectrum of BL Lacertae also depends on the Galactic absorption assumed. The amount of absorption due to molecular hydrogen is not directly measurable, making the estimation of the total absorption along our line of sight uncertain. The Galactic atomic hydrogen column density towards BL Lacertae is $N_{HI}$ = 1.75$\times$10$^{21}$\,cm$^{-2}$, as obtained by the HI4PI survey \citep{benbekhti16}. Approximately the same amount of Galactic atomic hydrogen column density has been previously estimated with the Leiden/Argentine/Bonn (LAB) survey \citep[1.73$\times$10$^{21}$\,cm$^{-2}$;][]{kalberla05}. However, observations of local interstellar CO have shown the presence of a Galactic molecular cloud towards the source \citep[e.g.,][]{bania91, liszt98}. The total hydrogen column density towards BL Lacertae ($N_{H,tot}$) is thus composed by the atomic hydrogen column density, $N_{HI}$, and the molecular column density, $N_{H_2}$. The estimation of the amount of $N_{H_2}$ reported in literature changes significantly depending on the X-ray satellites used, from 0.5$\times$10$^{21}$\,cm$^{-2}$ using the ASCA data \citep{madejski99} to 1.7$\times$10$^{21}$\,cm$^{-2}$ \citep{raiteri09} using the {\em XMM-Newton} data. 

According to \citet{liszt98}, the $\rm^{13}CO$ column density of the molecular cloud is (8.48 $\pm$ 0.78) $\times$10$^{14}$\,cm$^{-2}$. Assuming that the molecular hydrogen column density $N_{\rm H_2}$ is usually (1--2) $\times$$10^6$ times the $\rm^{13}CO$ one \citep[see e.g.,][]{liszt07}, the estimation of hydrogen column density due to the molecular cloud varies between 7.7$\times$10$^{20}$ and 18.6$\times10^{21}$\,cm$^{-2}$. This results in a total hydrogen column density towards BL Lacertae of (2.52--3.61)$\times$10$^{21}$ cm$^{-2}$. However, the CO component can provide only very approximately about the molecular component of the gas. Moreover, the value also depends on the ratio $N_{H_2} / N_{\rm ^{13}CO}$ and therefore to the corresponding uncertainties. Finally, a change of 14 per cent of the equivalent width of the H$_{2}$CO absorption lines along the line of sight of BL Lacertae in 2 yr has been reported by \citet{moore95}, suggesting possible variability of the molecular column density. The best way to estimate the value of $N_{H,tot}$ in BL Lacertae is to have an observation with a large number of counts, and covering a broad energy range. For this reason, the joint {\em NICER} and {\em NuSTAR} spectrum collected on 2020 October 11--12 is ideal for such a kind of study. Leaving the total Galactic absorption free to vary and using a broken power-law model, we obtain an $N_{H,tot}$ = (2.59\,$^{+0.14}_{-0.09}$) $\times$10$^{21}$ cm$^{-2}$. 

By applying the relationship between dust-emission-derived reddening E(B-V) and hydrogen column density, $N_{H_I}$ = 8.3 $\times$10$^{21}$\,cm$^{-2}$$\times$E(B-V), and using E(B-V) = 0.316 mag, we obtain 2.62$\times$10$^{21}$\,cm$^{-2}$, comparable to the value obtained by the {\em NICER} and {\em NuSTAR} joint fit.

Further X-ray observations of blazars in high-activity periods for long exposures and over a broad band energy range, as assured by the combination of {\em NICER} and {\em NuSTAR} observations, are fundamental to better characterize the X-ray spectrum of these sources and to perform searches for rapid X-ray variability. In this context, this study has shown that {\em NICER} can be also important for blazar science, even more if combined with the {\em NuSTAR} data at hard X-rays. 

\section*{Acknowledgements}

We thank the NuSTAR PI, Fiona Harrison, for approving the DDT request, and the NuSTAR SOC for carrying out the observation and the excellent support. 

\noindent We thank the NICER PI, Keith Gendreau, and the operation team for the rapid approval and execution of the DDT request. We thank the {\em Swift} team for making these observations possible, the PI, Brad Cenko, the duty scientists, and science planners. 

This research has made use of the XRT Data Analysis Software (XRTDAS). This work made use of data supplied by the UK Swift Science Data Centre at the University of Leicester. This research has made use of data obtained through the High Energy Astrophysics Science Archive Research Center Online Service, provided by the NASA/Goddard Space Flight Center. This research has made use of the NASA/IPAC Extragalactic Database (NED), which is operated by the Jet Propulsion Laboratory, California Institute of Technology, under contract with the National Aeronautics and Space Administration. We thank the second referee, Dirk Grupe, for constructive comments and suggestions.

\section*{Data Availability}

The data underlying this article are available in {\em Swift} Archive Download Portal at the UK {\em Swift} Science Data Centre (https://www.swift.ac.uk/swift\_portal/) and the HEASARC database (https://heasarc.gsfc.nasa.gov/docs/archive.html).

\label{lastpage}

\clearpage

\appendix

\section{Swift results}\label{Appendix1}

\subsection{{\em Swift}-XRT}

\begin{table*}
\caption{Log and fitting results of {\em Swift}-XRT observations of BL Lacertae using a power-law model with $N_{\rm H}$ fixed to 2.70$\times$10$^{21}$ cm$^{-2}$. Fluxes are corrected for the Galactic absorption. MJD refers to the start time of the {\em Swift} observation.} 
\label{A1}
\begin{center}
\begin{tabular}{ccccc}
\hline
\multicolumn{1}{c}{\textbf{MJD}} &
\multicolumn{1}{c}{\textbf{Date}} &
\multicolumn{1}{c}{\textbf{Net exposure time}} &
\multicolumn{1}{c}{\textbf{Photon index}}      &
\multicolumn{1}{c}{\textbf{Flux$_{\rm\,0.3-10\,keV}$}}  \\
\multicolumn{1}{c}{(UT)} &
\multicolumn{1}{c}{} &
\multicolumn{1}{c}{(s)} &
\multicolumn{1}{c}{($\Gamma_{\rm\,X}$)}  &
\multicolumn{1}{c}{(10$^{-11}$ erg cm$^{-2}$ s$^{-1}$)}  \\
\hline
59072.013301  &  2020-08-11 &   1144  &  1.95 $\pm$  0.15   &     2.36  $\pm$  0.26    \\ 
59072.979057  &  2020-08-11 &   2460  &  1.73 $\pm$  0.12   &     1.67  $\pm$  0.16    \\ 
59074.007524  &  2020-08-13 &   1089  &  1.78 $\pm$  0.18   &     1.49  $\pm$ 0.21     \\ 
59080.300875  &  2020-08-19 &    986  &  2.19 $\pm$  0.13   &     3.16  $\pm$  0.29    \\
59081.600112  &  2020-08-20 &    974  &  1.76 $\pm$  0.16   &     2.20  $\pm$  0.29    \\ 
59082.265103  &  2020-08-21 &    979  &  1.87 $\pm$  0.15   &     2.74  $\pm$  0.30    \\ 
59083.054394  &  2020-08-22 &    979  &  1.87 $\pm$  0.14   &     3.40  $\pm$  0.37    \\ 
59084.018388  &  2020-08-23 &    652  &  1.82 $\pm$  0.18   &     2.70  $\pm$  0.37    \\ 
59085.319328  &  2020-08-24 &   1199  &  1.95 $\pm$  0.15   &     2.45  $\pm$  0.27    \\ 
59086.086696  &  2020-08-25 &    854  &  2.08 $\pm$  0.17   &     2.32  $\pm$  0.27    \\ 
59087.077198  &  2020-08-26 &    814  &  1.90 $\pm$  0.17   &     2.41  $\pm$  0.30    \\ 
59088.073026  &  2020-08-27 &    697  &  1.88 $\pm$  0.19   &     2.18  $\pm$  0.32    \\  
59098.704090  &  2020-09-06 &    609  &  1.60 $\pm$  0.18   &     2.92  $\pm$  0.44    \\  
59104.806412  &  2020-09-12 &    922  &  1.74 $\pm$  0.18   &     1.96  $\pm$  0.31    \\  
59113.191735  &  2020-09-21 &   1983  &  2.20 $\pm$  0.10   &     3.27  $\pm$  0.22    \\ 
59114.367874  &  2020-09-22 &    864  &  2.22 $\pm$  0.18   &     2.42  $\pm$  0.15    \\  
59116.151464  &  2020-09-24 &    817  &  2.25 $\pm$  0.15   &     3.14  $\pm$  0.16    \\  
59117.420004  &  2020-09-25 &   1408  &  1.95 $\pm$  0.16   &     1.74  $\pm$  0.20    \\  
59124.730783  &  2020-10-02 &    984  &  2.48 $\pm$  0.11   &     9.99  $\pm$  0.70    \\  
59125.655695  &  2020-10-03 &    989  &  1.93 $\pm$  0.18   &     1.76  $\pm$  0.23    \\  
59126.597595  &  2020-10-04 &    135  &  2.59 $\pm$  0.50   &     1.90  $\pm$  0.74    \\  
59127.638321  &  2020-10-05 &    989  &  2.37 $\pm$  0.12   &    16.90  $\pm$  1.26    \\  
59128.182674  &  2020-10-06 &    953  &  2.58 $\pm$  0.02   &    65.28  $\pm$  1.16    \\  
59128.942192  &  2020-10-06 &   1426  &  2.63 $\pm$  0.08   &    17.14  $\pm$  0.87    \\  
59129.906850  &  2020-10-07 &    237  &  2.53 $\pm$  0.28   &     3.38  $\pm$  0.56    \\  
59130.907624  &  2020-10-08 &   1558  &  2.32 $\pm$  0.10   &     3.60  $\pm$  0.25    \\ 
59131.900956  &  2020-10-09 &   1483  &  2.43 $\pm$  0.12   &     6.83  $\pm$  0.53    \\  
59132.891488  &  2020-10-10 &   1463  &  2.60 $\pm$  0.10   &     5.35  $\pm$  0.31    \\  
59133.850977  &  2020-10-11 &   2934  &  1.64 $\pm$  0.12   &     1.19  $\pm$  0.12    \\   
59134.891522  &  2020-10-12 &    654  &  2.19 $\pm$  0.23   &     1.70  $\pm$  0.26    \\   
59135.835244  &  2020-10-13 &   2307  &  1.77 $\pm$  0.12   &     1.72  $\pm$  0.15    \\   
59137.660548  &  2020-10-14 &   1051  &  1.50 $\pm$  0.25   &     1.01  $\pm$  0.19    \\   
59138.269685  &  2020-10-15 &   1628  &  1.35 $\pm$  0.17   &     1.21  $\pm$  0.18    \\ 
\hline                           
\end{tabular}                   
\end{center}
\end{table*}

\begin{figure*}
 \begin{center}
 \includegraphics[width=0.75\textwidth]{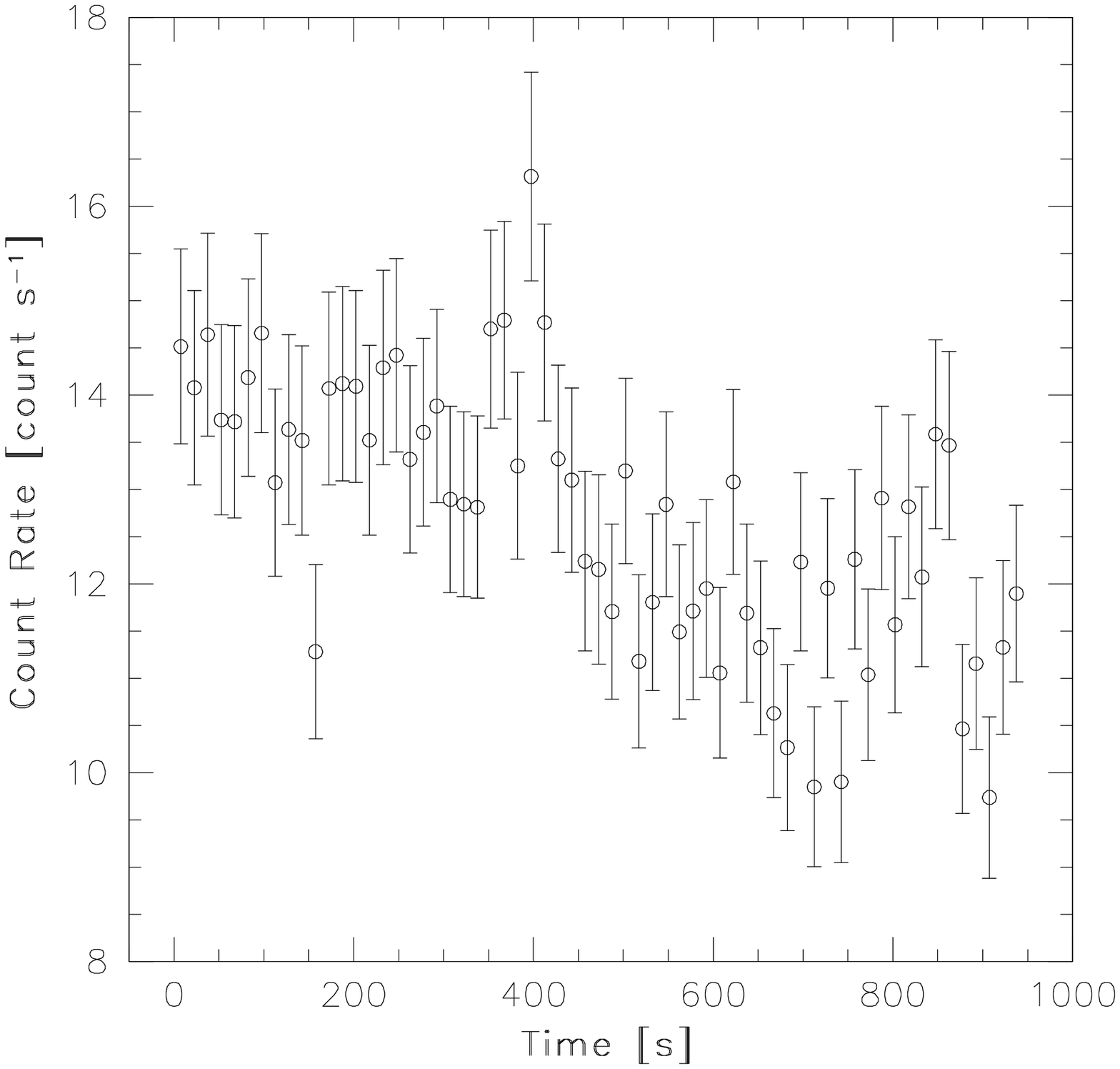}
 \caption{{\em Swift}-XRT light curve of BL Lacertae collected on 2020 October 6 shown in terms of count rate using 15 s time bins.}
 \label{A1bis}
 \end{center}
 \end{figure*}

\clearpage

\subsection{{\em Swift}-UVOT}
\begin{table*}
\caption{Observed magnitude (i.e. not corrected for Galactic extinction and the contribution of the host galaxy) of BL Lacertae obtained by {\em Swift}-UVOT. MJD refers to the start time of the {\em Swift} observation.}
\label{UVOT}
\begin{center}
\begin{tabular}{cccccccc}
\hline
\multicolumn{1}{c}{\textbf{MJD}}       &
\multicolumn{1}{c}{\textbf{Date (UT)}} &
\multicolumn{1}{c}{\textbf{$v$}}       &
\multicolumn{1}{c}{\textbf{$b$}}       &
\multicolumn{1}{c}{\textbf{$u$}}       &
\multicolumn{1}{c}{\textbf{$w1$}}      &
\multicolumn{1}{c}{\textbf{$m2$}}      &
\multicolumn{1}{c}{\textbf{$w2$}}      \\
\hline
59072.013301  &  2020-08-11 &   12.75 $\pm$   0.04 & 13.56 $\pm$   0.05 & 13.08 $\pm$  0.05 & 13.59 $\pm$ 0.06 & 14.23  $\pm$  0.06 &  14.27 $\pm$  0.06  \\ 
59072.979057  &  2020-08-11 &   13.21 $\pm$   0.03 & 14.01 $\pm$   0.04 & 13.54 $\pm$  0.05 & 14.10 $\pm$ 0.06 & 14.82  $\pm$  0.06 &  14.82 $\pm$  0.06  \\ 
59074.007524  &  2020-08-13 &   13.02 $\pm$   0.04 & 13.82 $\pm$   0.05 & 13.38 $\pm$  0.05 & 13.96 $\pm$ 0.06 & 14.61  $\pm$  0.07 &  14.63 $\pm$  0.06  \\ 
59080.300875  &  2020-08-19 &   12.78 $\pm$   0.03 & 13.53 $\pm$   0.05 & 13.05 $\pm$  0.05 & 13.61 $\pm$ 0.06 & 14.24  $\pm$  0.06 &  14.24 $\pm$  0.06  \\
59081.600112  &  2020-08-20 &   12.58 $\pm$   0.04 & 13.34 $\pm$   0.05 & 12.86 $\pm$  0.05 & 13.46 $\pm$ 0.05 & 14.15  $\pm$  0.15 &  14.12 $\pm$  0.06  \\ 
59082.265103  &  2020-08-21 &          -           & 13.43 $\pm$   0.04 & 12.95 $\pm$  0.05 & 13.53 $\pm$ 0.05 &        -           &  14.17 $\pm$  0.06  \\ 
59083.054394  &  2020-08-22 &          -           & 13.34 $\pm$   0.04 & 12.87 $\pm$  0.05 & 13.42 $\pm$ 0.05 &        -           &  14.10 $\pm$  0.06  \\ 
59084.018388  &  2020-08-23 &   12.54 $\pm$   0.04 & 13.31 $\pm$   0.05 & 12.83 $\pm$  0.05 & 13.36 $\pm$ 0.06 & 14.06  $\pm$  0.06 &  14.08 $\pm$  0.06  \\ 
59085.319328  &  2020-08-24 &   12.70 $\pm$   0.04 & 13.49 $\pm$   0.04 & 13.01 $\pm$  0.05 & 13.56 $\pm$ 0.06 & 14.23  $\pm$  0.06 &  14.23 $\pm$  0.06  \\ 
59086.086696  &  2020-08-25 &   12.75 $\pm$   0.04 & 13.54 $\pm$   0.05 & 13.01 $\pm$  0.05 & 13.59 $\pm$ 0.06 & 14.25  $\pm$  0.06 &  14.21 $\pm$  0.06  \\ 
59087.077198  &  2020-08-26 &   12.89 $\pm$   0.04 & 13.68 $\pm$   0.05 & 13.16 $\pm$  0.05 & 13.74 $\pm$ 0.06 & 14.31  $\pm$  0.07 &  14.33 $\pm$  0.06  \\ 
59088.073026  &  2020-08-27 &   13.42 $\pm$   0.04 & 14.23 $\pm$   0.05 & 13.84 $\pm$  0.05 & 14.42 $\pm$ 0.07 & 15.07  $\pm$  0.07 &  15.14 $\pm$  0.07  \\ 
59098.704090  &  2020-09-06 &   13.04 $\pm$   0.05 & 13.85 $\pm$   0.05 & 13.42 $\pm$  0.05 & 14.05 $\pm$ 0.05 &        -           &  14.71 $\pm$  0.06  \\ 
59104.806412  &  2020-09-12 &   13.45 $\pm$   0.05 & 14.23 $\pm$   0.05 & 13.83 $\pm$  0.05 & 14.49 $\pm$ 0.05 & 15.16  $\pm$  0.07 &  15.22 $\pm$  0.07  \\ 
59113.191735  &  2020-09-21 &   13.14 $\pm$   0.05 & 13.89 $\pm$   0.04 & 13.41 $\pm$  0.05 & 13.95 $\pm$ 0.06 & 14.62  $\pm$  0.06 &  14.61 $\pm$  0.06  \\ 
59114.367874  &  2020-09-22 &   13.00 $\pm$   0.04 & 13.71 $\pm$   0.05 & 13.23 $\pm$  0.05 & 13.78 $\pm$ 0.06 & 14.43  $\pm$  0.07 &  14.48 $\pm$  0.06  \\ 
59116.151464  &  2020-09-24 &   13.02 $\pm$   0.04 & 13.72 $\pm$   0.05 & 13.22 $\pm$  0.05 & 13.78 $\pm$ 0.06 & 14.31  $\pm$  0.07 &  14.39 $\pm$  0.06  \\ 
59117.420004  &  2020-09-25 &   13.05 $\pm$   0.04 & 13.84 $\pm$   0.04 & 13.34 $\pm$  0.05 & 13.96 $\pm$ 0.06 & 14.60  $\pm$  0.06 &  14.68 $\pm$  0.06  \\ 
59124.730783  &  2020-10-02 &   12.72 $\pm$   0.04 & 13.46 $\pm$   0.04 & 12.97 $\pm$  0.05 & 13.51 $\pm$ 0.06 & 14.19  $\pm$  0.11 &  14.19 $\pm$  0.06  \\ 
59125.655695  &  2020-10-03 &         -            & 13.87 $\pm$   0.04 & 13.42 $\pm$  0.05 & 14.01 $\pm$ 0.06 &        -           &  14.68 $\pm$  0.06  \\ 
59126.597595  &  2020-10-04 &   13.23 $\pm$   0.07 & 13.99 $\pm$   0.07 & 13.53 $\pm$  0.07 & 14.07 $\pm$ 0.09 & 14.93  $\pm$  0.15 &  14.79 $\pm$  0.09  \\ 
59127.638321  &  2020-10-05 &   12.40 $\pm$   0.04 & 13.14 $\pm$   0.05 & 12.56 $\pm$  0.05 & 13.08 $\pm$ 0.06 & 13.70  $\pm$  0.06 &  13.67 $\pm$  0.06  \\ 
59128.182674  &  2020-10-06 &   12.62 $\pm$   0.04 & 13.30 $\pm$   0.05 & 12.72 $\pm$  0.05 & 13.19 $\pm$ 0.06 & 13.73  $\pm$  0.06 &  13.76 $\pm$  0.06  \\ 
59128.942192  &  2020-10-06 &   12.85 $\pm$   0.04 & 13.62 $\pm$   0.05 & 13.10 $\pm$  0.05 & 13.68 $\pm$ 0.05 & 14.21  $\pm$  0.06 &  14.20 $\pm$  0.06  \\ 
59129.906850  &  2020-10-07 &         -            &       -            & 13.32 $\pm$  0.06 & 13.89 $\pm$ 0.06 &        -           &        -            \\ 
59130.907624  &  2020-10-08 &   13.22 $\pm$   0.04 & 13.99 $\pm$   0.04 & 13.47 $\pm$  0.05 & 14.03 $\pm$ 0.06 & 14.65  $\pm$  0.07 &  14.66 $\pm$  0.06  \\ 
59131.900956  &  2020-10-09 &   12.91 $\pm$   0.04 & 13.61 $\pm$   0.04 & 13.14 $\pm$  0.05 & 13.67 $\pm$ 0.06 & 14.38  $\pm$  0.07 &  14.31 $\pm$  0.06  \\  
59132.891488  &  2020-10-10 &   13.00 $\pm$   0.04 & 13.74 $\pm$   0.04 & 13.19 $\pm$  0.05 & 13.74 $\pm$ 0.05 & 14.34  $\pm$  0.06 &  14.36 $\pm$  0.06  \\ 
59133.850977  &  2020-10-11 &   13.32 $\pm$   0.04 & 14.10 $\pm$   0.04 & 13.63 $\pm$  0.05 & 14.22 $\pm$ 0.06 & 14.89  $\pm$  0.06 &  14.92 $\pm$  0.06  \\ 
59134.891522  &  2020-10-12 &         -            & 14.01 $\pm$   0.06 & 13.54 $\pm$  0.05 & 14.11 $\pm$ 0.06 &        -           &  14.75 $\pm$  0.06  \\ 
59135.835244  &  2020-10-13 &   13.12 $\pm$   0.04 & 13.95 $\pm$   0.04 & 13.46 $\pm$  0.05 & 14.06 $\pm$ 0.06 & 14.69  $\pm$  0.06 &  14.77 $\pm$  0.06  \\ 
59137.660548  &  2020-10-14 &   13.48 $\pm$   0.04 & 14.32 $\pm$   0.05 & 13.87 $\pm$  0.05 & 14.50 $\pm$ 0.06 & 15.13  $\pm$  0.07 &  15.16 $\pm$  0.06  \\ 
59138.269685  &  2020-10-15 &   13.61 $\pm$   0.04 & 14.42 $\pm$   0.05 & 13.92 $\pm$  0.05 & 14.55 $\pm$ 0.06 & 15.30  $\pm$  0.06 &  15.27 $\pm$  0.07  \\ 
\hline
\end{tabular}
\end{center}
\end{table*} 

\clearpage

\section{NuSTAR and NICER results}\label{Appendix2}
\begin{figure*}
 \begin{center}
 \includegraphics[width=0.75\textwidth]{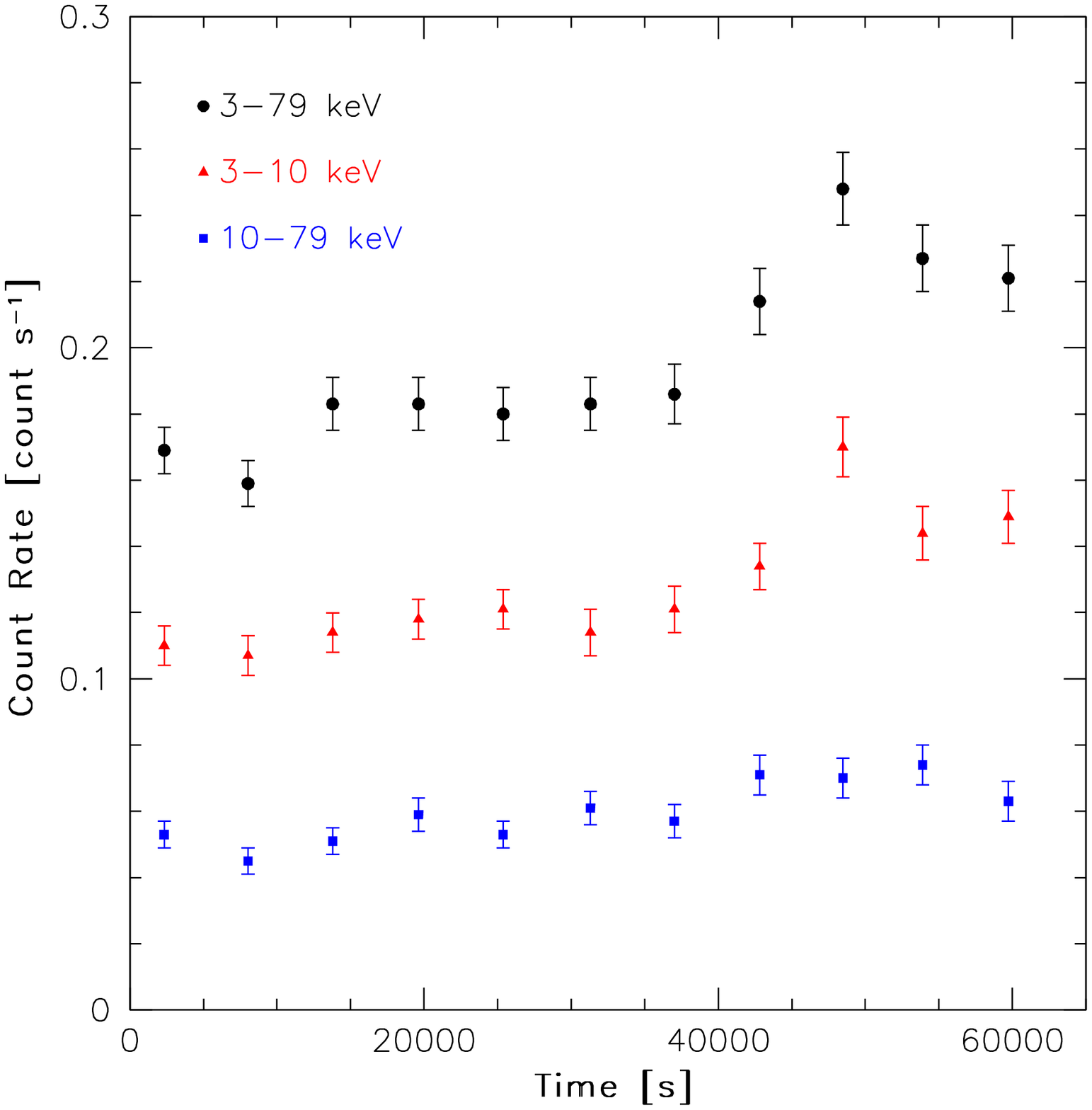}
 \caption{Count rate per orbit for {\em NuSTAR} as seen by the FPMA module in the 3--79 keV (black dots), 3--10 keV (red triangles), and 10--79 keV (blue squares) energy range, respectively.}
 \label{B1}
 \end{center}
 \end{figure*}
 
\begin{figure*}
 \begin{center}
 \includegraphics[width=0.75\textwidth]{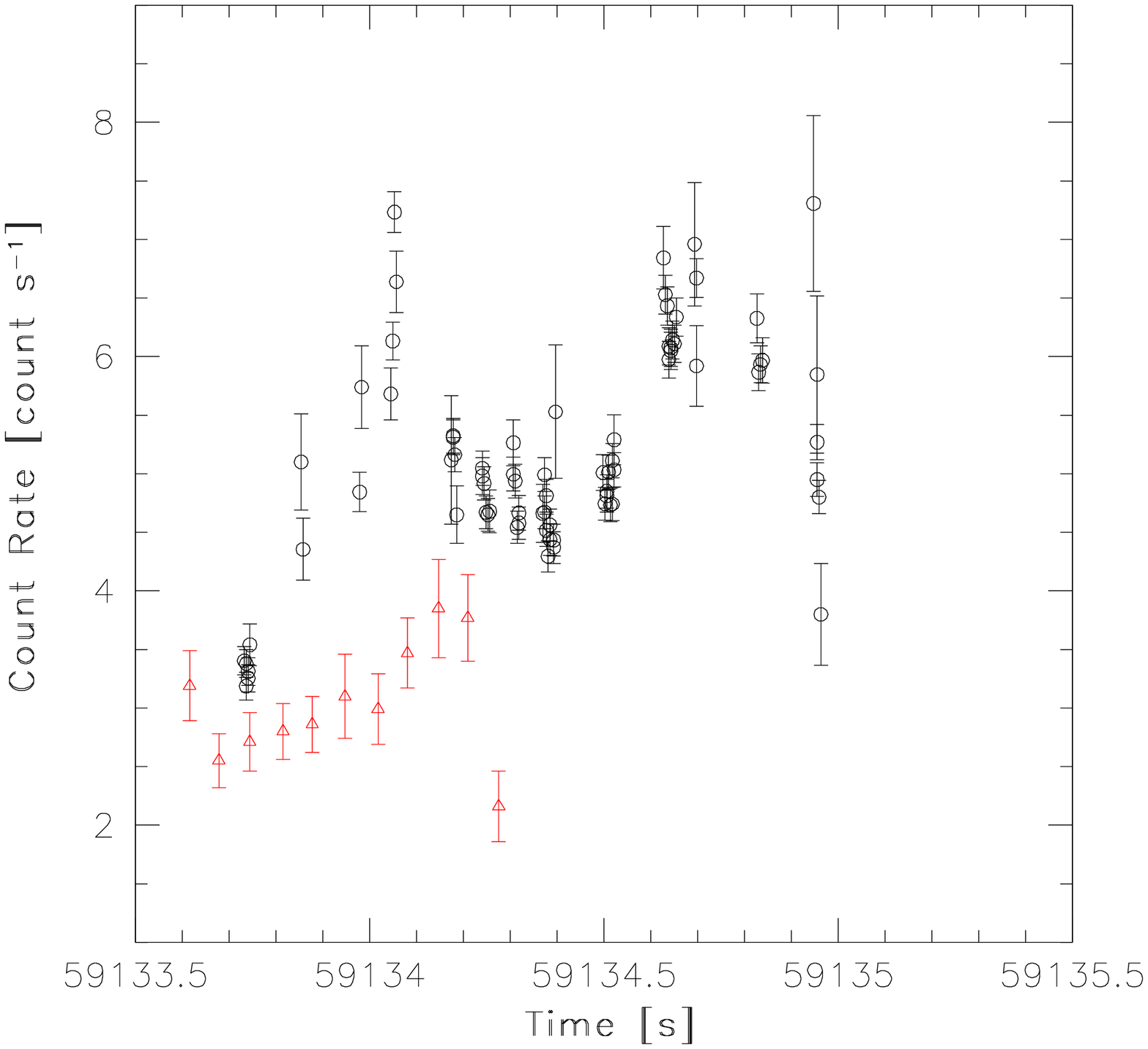}
 \caption{Comparison of the count rate obtained for {\em NICER} in the 0.4--5.0 keV energy range (black open circles) and {\em NuSTAR} (red filled triangles) in the 3--79 keV energy range (black dots) during 2020 October 11--12.}
 \label{B2}
 \end{center}
 \end{figure*} 

\end{document}